\documentclass[apj]{emulateapj}

\usepackage{multirow}
\usepackage{longtable}
\usepackage{setspace}
\usepackage{amsmath}
\usepackage{amssymb,hyperref}
\usepackage{xcolor}

\def\erg{{\rm\thinspace erg}}

\def\km{{\rm\thinspace km}}

\def\Mpc{{\rm\thinspace Mpc}}

\def\s{{\rm\thinspace s}}

\def\ergps{\hbox{$\erg\s^{-1}\,$}}

\def\kmpspMpc{\hbox{$\km\s^{-1}\Mpc^{-1}\,$}}

\def\bootes{{Bo\"{o}tes~}}

\shorttitle{Evolution of AGN hosts to $z\sim 1.4$ }
\shortauthors{A.D.~Goulding et al.}

\begin{document}

\title{Tracing the evolution of Active Galactic Nuclei \\
  host galaxies over the last 9 Gyrs of Cosmic time}

\author{A. D. Goulding\altaffilmark{1}, W.~R. Forman\altaffilmark{1},
  R.~C. Hickox\altaffilmark{2}, C. Jones\altaffilmark{1},
  S.~S. Murray\altaffilmark{1,3}, \\
  A. Paggi\altaffilmark{1}, M.~L.~N. Ashby\altaffilmark{1},
  A.~L.~Coil\altaffilmark{4}, M.~C. Cooper\altaffilmark{5},
  J.-S. Huang\altaffilmark{1}, R. Kraft\altaffilmark{1}, \\
  J.~A. Newman\altaffilmark{6}, B.~J. Weiner\altaffilmark{7} and
  S.~P. Willner\altaffilmark{1}}

\email{agoulding@cfa.harvard.edu}
\altaffiltext{1}{Harvard-Smithsonian Center for Astrophysics, 60
  Garden St., Cambridge, MA 02138, USA}
\altaffiltext{2}{Department of Physics and Astronomy, Dartmouth
 College, Hanover, NH 03755, USA}
\altaffiltext{3}{Department of Physics and Astronomy, Johns Hopkins
 University, 3400 North Charles Street, Baltimore, MD 21218, USA}
\altaffiltext{4}{Department of Physics, Center for Astrophysics and Space Sciences,
  University of California at San Diego, 9500 Gilman Dr., La Jolla,
  San Diego, CA 92093}
\altaffiltext{5}{Center for Galaxy Evolution, Department of Physics
  and Astronomy, University of California, Irvine, 4129 Frederick
  Reines Hall, Irvine, CA 92697, USA}
\altaffiltext{6}{Department of Physics and Astronomy, University of Pittsburgh, 3941
  O'Hara Street, Pittsburgh, PA 15260, USA}
\altaffiltext{7}{Steward Observatory, 933 North Cherry Street,
  University of Arizona, Tucson, AZ 85721}

\begin{abstract}
  We present the results of a combined galaxy population analysis for
  the host galaxies of active galactic nuclei (AGN) identified at
  $0<z<1.4$ within the SDSS, \bootes and DEEP2 surveys. We identified
  AGN in a uniform and unbiased manner at X-ray, infrared and radio
  wavelengths. Supermassive black holes undergoing
  radiatively-efficient accretion (detected as X-ray and/or infrared
  AGN) appear to be hosted in a separate and distinct galaxy
  population than AGN undergoing powerful mechanically dominated
  accretion (radio AGN). Consistent with some previous studies,
  radiatively efficient AGN appear to be preferentially hosted in
  modest star-forming galaxies, with little dependence on AGN or
  galaxy luminosity. AGN exhibiting radio-emitting jets due to
  mechanically-dominated accretion are almost exclusively observed in
  massive, passive galaxies. Crucially, we now provide strong evidence
  that the observed host-galaxy trends are independent of redshift. In
  particular, these different accretion-mode AGN have remained as
  separate galaxy populations throughout the last 9 Gyr. Furthermore,
  it appears that galaxies hosting AGN have evolved along the same
  path as galaxies that are not hosting AGN with little evidence for
  distinctly separate evolution.
\end{abstract}

\keywords{galaxies: active; galaxies: evolution; galaxies: statistics;
  AGN; infrared: galaxies; radio continuum: galaxies; X-rays:
  galaxies; surveys}

\section{Introduction} \label{sec:intro}

Accreting supermassive black holes (BHs) in the form of active
galactic nuclei (AGN) are capable of releasing enormous quantities of
energy over their lifetimes, often comparable to the binding energy of
their host galaxies ($> 10^{61}$~ergs; for a review see
\citealt{dma12}). In light of this, many theoretical and
semi-analytical galaxy-evolution simulations now incorporate AGN
feedback processes as a form of self regulation for BH growth and
star formation (SF). It is predicted that major mergers between
gas-rich galaxies drive nuclear inflows, which trigger powerful
starbursts and fuel the growth of central BHs
(\citealt{dimatteo05,hopkins07,hopkins08b,menci08}). In turn, these
models often reproduce key observables such as the bulge--BH mass
scaling relations (e.g., \citealt{magorrian98,tremaine02}) and the
observed luminosity functions of quasars and normal galaxies (e.g.,
\citealt{Richards06,ross12}).

Despite the success of current models, it is now becoming clear that
galaxy--galaxy mergers, which are required by some of the cosmological
simulations, are relatively rare events, and may not be driving the
bulk of galaxy and AGN co-evolution (for a recent review of AGN/galaxy
co-evolution see \citealt{kormendy13}). Indeed, dedicated
multi-wavelength (optical; X-ray; far-infrared) studies are providing
mounting evidence that the majority of AGN--galaxy co-evolution is
dominated by more secular processes (i.e., due to material accreted
directly from the host galaxy). Analysis of the distribution of AGN
morphologies and their position in galaxy color versus absolute
magnitude space shows that at $z\sim 1$--2, AGN hosts are
predominantly star forming and disk dominated (`blue cloud') galaxies
and not merging systems (e.g.,
\citealt{georgakakis09,schawinski11,cisternas11,mullaney12,rosario12}). There
have been suggestions that relatively isolated X-ray AGN may represent
a specific population of galaxies that are undergoing a transition in
their stellar and morphological properties (e.g.,
\citealt{nandra07,coil09,hickox09,xue10}), and this transition is due
to the central AGN. The most extreme cases of this transition
population may be due to the presence of a luminous quasar with
accretion in excess of 10\% of the Eddington rate. A quasar phase is
expected to rapidly truncate on-going star formation by driving out
the available cool gas supply in the form of a disk wind that is
radiatively efficient (see \citealt{fabian12} for a review). Such a
rapid BH growth stage may (in-part) provide a feedback mechanism to
evolve blue galaxies onto the red sequence (e.g.,
\citealt{hopkins10}).

Unlike radiatively-efficient AGN and quasars that are ubiquitously
detected in optical/IR/X-ray surveys, radio AGN, which accrete matter
through advection dominated processes, are almost exclusively hosted
in old stellar population (`red sequence') massive spheroidal galaxies
(e.g., \citealt{best05b,kauffmann08}). These red-sequence galaxies are
expected to have evolved through catastrophic interactions such as
major mergers. The most powerful radio sources are often hosted in
massive spheroidal systems cocooned within hot atmospheres, such as
those at the centers of galaxy clusters and dense groups. The
mechanical energy released by the central radio source is injected
back into the interstellar and intracluster medium (so called
radio-mode feedback), reducing the production of new stars in these
massive (gas rich) systems. Tracing the hosts of these two
fundamentally different accretion mode (radiatively-efficient versus
advection-dominated) AGN populations during the epoch where the red
sequence of galaxies is forming \citep{bell04,borch06,faber07}
may provide valuable insight into galaxy evolution and AGN feedback.

Wide field spectroscopic surveys that include tens to hundreds of
thousands of extragalactic sources have shown that the galaxy
luminosity and stellar mass functions evolve strongly from $z \sim
1.5$ to the present day with the main build-up of the red sequence
occurring since $z \sim 1$
\citep{bell04,borch06,cooper06,faber07,moustakas13}. Clearly, $z \sim
1$ is an important epoch to study the interaction between growing
supermassive BHs and their hosts. The accurate identification and
selection of relatively distant galaxies and AGN requires robust
source redshift measurements. Indeed, to constrain the color
distributions and evolution of the AGN hosts, accurate spectroscopic
redshift measurements are required to break the degeneracy between
inferred properties (such as star formation history, stellar mass and
color) and redshift that is inherent in photometric redshift
determinations, particularly for surveys with limited multi-narrowband
photometric coverage.

In the nearby Universe ($z < 0.2$), surveys such as the Sloan Digital
Sky Survey and 2dF Galaxy Redshift Survey have provided an
unprecedented wealth of spectroscopic data across $\sim
10,000$~deg$^2$ (in extragalactic sky regions), which has been used
effectively to identify and understand the optical, radio and to some
extent, the X-ray and infrared (IR) properties of AGN and galaxies
locally (e.g.,
\citealt{kauff03b,heckman04,lamassa09,lamassa12,shao13}). However,
while useful for selecting AGN in nearby galaxies (to $z<0.2$), due to
brightness limitations (e.g., including only galaxies with $r < 17.77$
in the SDSS), these surveys sample only the most luminous and massive
systems at higher redshifts ($z > 0.2$). Furthermore, it has now been
widely established that no single waveband can accurately provide a
complete sample of AGN due primarily to: (1) absorption/obscuration of
the central BH; (2) contamination by the host galaxy (e.g.,
\citealt{hickox09,hopkins09,donley10,goulding12a,mendez13}); and (3)
individual wavebands preferentially selecting particular accretion
mechanisms (i.e., radiative versus advection dominated flows).

To trace the evolution of the typical population of AGN and their
hosts to $z \sim 1.4$ within similar and unbiased volume slices, we
require a combination of relatively wide-field, sensitive
multi-wavelength (optical; radio; X-ray; IR) surveys with complete
homogeneous spectroscopic coverage. Hence, here we harness the
extensive datasets in three wide and blank field surveys with complete
homogeneous optical spectroscopic coverage: the Sloan Digital Sky
Survey ($z$$\sim$0--0.3; \citealt{sdss_tech}), Bo\"{o}tes
($z$$\sim$0.25--0.8; \citealt{murray05,kochanek12}), and DEEP2
($z$$\sim$0.7--1.4; \citealt{davis03,newman13}). The redshift and
galaxy property information established using these surveys combined
with the AGN identified using the multi-wavelength data provides one
of the most complete views of AGN activity and the host galaxy
interaction at $z < 1.4$. Here we explore:
\begin{enumerate}
\item the multi-wavelength incidence of AGN activity in optically
  selected galaxies; \vspace{-0.21cm}

\item the distribution and median rest-frame optical colors of AGN
  host galaxies; \vspace{-0.21cm}

\item the (non-)evolution of AGN host galaxy colors throughout the
  last 9 Gyrs.
\end{enumerate}

Section 2 introduces the three survey fields used in this work and
describe how our galaxy sample was constructed. The full description
of the parent galaxy samples and how these are matched for the
differing detection methods and sensitivity limits of the individual
fields are provided in the Appendix. Section 3 presents the
methodology for building a obscuration unbiased sample of AGN
identified in the IR, X-ray and radio domains.  Section 4 discusses
the general host galaxy properties of radio, X-ray and IR detected AGN
(i.e., advection dominated versus radiatively efficient AGN), and we
show that there appears to be little evidence for evolution in the
galaxy colors for individual AGN populations in the last 9 Gyrs. In
Section 5 outlines the impact of our findings on AGN feedback
models. Finally, Section 6 summarizes our findings. Throughout the
manuscript we adopt a standard flat $\Lambda$CDM cosmology with $H_0 =
71\kmpspMpc$ and $\Omega_M = 0.3$.

\begin{figure*}[t]
\includegraphics[width=\textwidth]{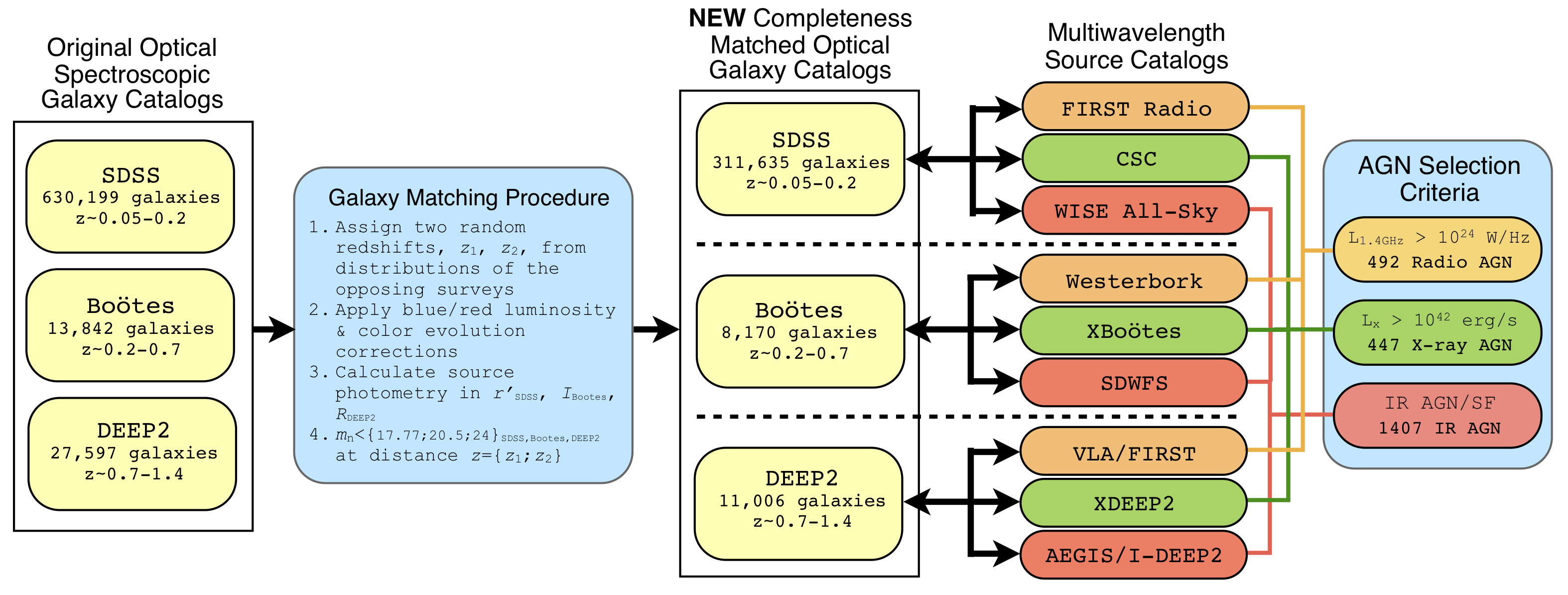}
\caption{A schematic flow diagram highlighting the main steps in the
  construction of our relatively unbiased, optical-property matched
  galaxy sample derived from the main SDSS-DR7, \bootes and DEEP2
  optical spectroscopic catalogs. Furthermore, we connect our matched
  galaxy sample, in each survey, to the appropriate multi-wavelength
  catalogs to identify those galaxies hosting luminous AGN.}
\label{fig:flowdiag}
\end{figure*}

\section{Galaxy Sample Selection} \label{sec:sample}

Our parent galaxy sample, defined at $z\sim0.05$--1.4, is constructed
from a combination of galaxies identified in the Seventh Data Release
of the Sloan Digital Sky Survey (hereafter, SDSS;
\citealt{sdss_tech,sdss_dr7}) at $0.05 < z < 0.2$, the NOAO Deep Wide
Field Survey \bootes field \citep{murray05,kochanek12} at $0.2 < z <
0.7$, and the DEEP Extragalactic Evolutionary Probe 2 (DEEP2) Galaxy
Redshift Survey fields \citep{davis03,newman13} at $0.7 < z <
1.4$. One of the primary aims of this experiment is to construct an
unbiased like-for-like galaxy sample from which to explore the
properties of galaxies hosting luminous AGN; i.e., the aim is to
construct a large galaxy sample selected consistently in all three
surveys, irrespective of the individual galaxy redshift, mass, color,
luminosity, and rest-frame flux limit of the survey. Hence, given that
the SDSS, Bo\"{o}tes, and DEEP2 surveys have different rest-frame
photometric and spectroscopic selection techniques, we cannot simply
use the full galaxy catalogs. The Appendix material provides a
thorough explanation of our sample selection technique to match the
galaxies identified in the three surveys according to passive
evolution corrected luminosity and spectral color across the entire
redshift range explored here. Briefly, we built upon the previous
methodology of \cite{blanton06} and performed photometric
$K$-corrections using the publicly available {\tt C} and {\tt IDL}
tool, {\sc kcorrect} (v4.2)\footnote{The current version of {\sc
    kcorrect} is available at
  \url{http://howdy.physics.nyu.edu/index.php/Kcorrect}}. These
measurements were then matched to the initial galaxy catalogs based on
their rest frame optical measurements in accordance with the selection
methodology of the surveys. We applied color and evolution corrections
to each source and assigned two random redshifts drawn from the
redshift distributions of the other two surveys. Those objects that
are included in our final galaxy sample are sufficiently luminous
(intrinsically) to be detected in all three parent surveys independent
of their specific observed redshift. Fig.\ref{fig:flowdiag} provides a
schematic flow diagram of the main processing steps and methodology
for the sample selection procedure.

Table~\ref{tbl:survey} provides the source statistics, photometric
limits, and redshift breakdowns for our color- and luminosity-matched
galaxy samples across the three surveys. Those galaxies included in
the final matched sample have their templates deprojected into SDSS
$u$ and $g$ filters in AB magnitudes. To aid comparison with previous
studies (e.g.,
\citealt{blanton03,blanton03c,blanton06,blanton07,kauffmann03b,hickox09}),
the SDSS filters are shifted blue-ward to $z=0.1$ and are defined as
$^{0.1}u$ and $^{0.1}g$, respectively. Absolute magnitudes, including
the passive evolution corrections, are calculated for each source. As
is standard with {\sc kcorrect}, the absolute magnitudes are
calculated assuming $h_0=1.0$, which we correct to $h_0 = 0.71$ to fit
our assumed cosmology.

\begin{figure}[t]
\hspace{-0.5cm}
\includegraphics[width=\linewidth]{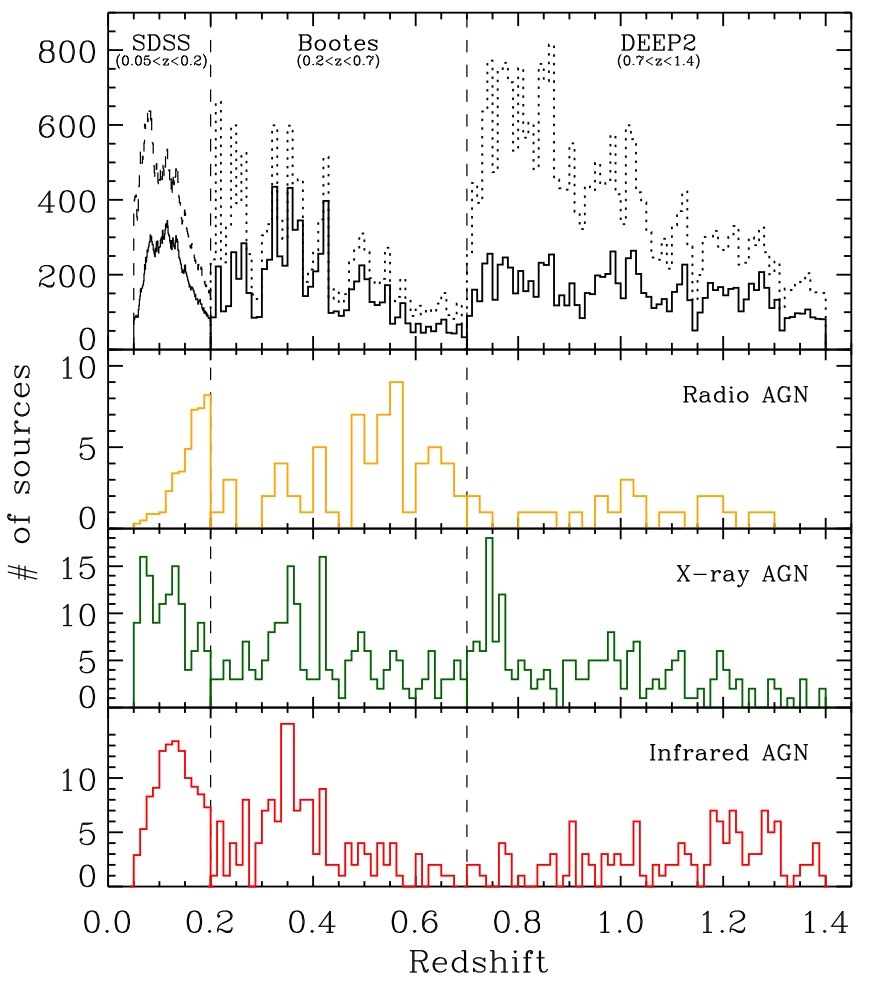}
\caption{{\it Top panel:} Spectral redshift distributions for the full
  parent catalogs (dashed-line; SDSS-MPA-JHU; Kochanek et al. 2012;
  Newman et al. 2013) as well as the selection and evolution matched
  galaxy samples (solid line). The final matched galaxy samples
  contain 311,685, 8,170 and 11,006 sources in SDSS, \bootes and
  DEEP2, respectively. {\it Lower panels: } Spectral redshift
  distributions for AGN identified at radio (yellow-solid), X-ray
  (green-solid) and infrared (red-solid) wavelengths in the survey
  fields. The histograms of the radio and IR AGN in the SDSS are
  scaled by a factor of 1/10 for presentation purposes.}
\label{fig:z_hists}
\end{figure}

The top panel of Fig.~\ref{fig:z_hists} shows the redshift
distributions for our matched galaxy sample and for comparison
purposes, the redshift distributions for the full galaxy samples in
the SDSS, Bo\"{o}tes, and DEEP2 surveys. At $z<0.7$, we find broadly
similar redshift distributions between the main galaxy samples and our
matched sample, suggesting that the matched sample is not overly
biased from the main population. At $0.75 < z < 0.85$, there is an
abundant population of galaxies that are present in the main DEEP2
survey but are not selected in our matched sample. These are
low-luminosity $^{0.1}M_{g} > -19.5$ blue galaxies that are not
selected in the matched sample due to the redder rest frame wavelength
selection in \bootes and the SDSS. The redshift distribution for the
DEEP2 property-matched galaxies is consistent with that expected when
matched to \bootes and the SDSS. In the following sections, we
identify galaxies that host AGN and use these samples to place
constraints on the AGN host galaxy properties.

\begin{table*}
\footnotesize
\begin{center}
\setlength{\tabcolsep}{1.5mm}
\caption{Survey Properties\label{tbl:survey}}
\begin{tabular}{lccccccccccc}
  \tableline\tableline
  \multicolumn{12}{c}{} \\[-0.5ex]
  \multicolumn{1}{c}{Survey} &
  \multicolumn{1}{c}{$z$} &
  \multicolumn{1}{c}{$m_{\rm limit}$} &
  \multicolumn{1}{c}{Gals with Spec-$z$} &
  \multicolumn{1}{c}{Completeness} &
  \multicolumn{4}{c}{Area (deg$^2$)} &
  \multicolumn{3}{c}{\# AGN} \\
  \multicolumn{1}{c}{} &
  \multicolumn{1}{c}{} &
  \multicolumn{1}{c}{(AB mag)} &
  \multicolumn{1}{c}{in $z$-range} &
  \multicolumn{1}{c}{Matched galaxies} &
  \multicolumn{1}{c}{Spec} &
  \multicolumn{1}{c}{X-ray} &
  \multicolumn{1}{c}{IR} &
  \multicolumn{1}{c}{Radio} &
  \multicolumn{1}{c}{X-ray} &
  \multicolumn{1}{c}{IR} &
  \multicolumn{1}{c}{Radio} \\[1ex]
  \multicolumn{1}{c}{(1)} &
  \multicolumn{1}{c}{(2)} &
  \multicolumn{1}{c}{(3)} &
  \multicolumn{1}{c}{(4)} &
  \multicolumn{1}{c}{(5)} &
  \multicolumn{4}{c}{(6)} &
  \multicolumn{3}{c}{(7)} \\[1ex]
  \tableline
  \\
  SDSS & 0.05--0.2 & $r<17.77$ & 630,199 & 311,635 ($\sim$49\%) & 8032 & 130 & 8032 & 7550 & 73 (122$^{\dagger}$) & 1121 & 406 \\ 
  \bootes & 0.2--0.7 & $I<20.5$ & 13,842 & 8,170 ($\sim$59\%) & 7.74 & 7.64 & 7.30 & 5.83 & 196 & 151 & 62 \\ 
  DEEP2 & 0.7--1.4 & $R<24.1$ & 27,597 & 11,006 ($\sim$40\%) & 2.78 & 2.66 & 1.64 & 2.78 & 178 & 135 & 24 \\
  \\
  \tableline\tableline
\end{tabular}
\end{center}
{\bf Notes:-}
$(1)$ Survey field; DEEP2$=$F1$+$F2$+$F3$+$F4.
$(2)$ Redshift range.
$(3)$ Brightness limit of the complete optical spectroscopic samples.
$(4)$ Number of sources with spectroscopic redshifts within the survey
redshift range.
$(5)$ Number of sources in the completeness limit matched sample.
$(6)$ Spatial area of the spectroscopic survey in square-degrees, and the contiguous
area covered by {\it Chandra} X-ray, WISE or {\it Spitzer} mid-IR and
VLA or Westerbork radio photometry. 
$(7)$ Number of AGN detected in the X-ray, IR and radio from the
matched source sample. $\dagger$: X-ray AGN in SDSS with $L_{X} > 10^{41}$\ergps.
\vspace{0.5cm}
\end{table*}

\section{AGN Sample Selection} \label{sec:AGN}

Many focused and blank field studies have shown that no one waveband
can identify the full population of AGN (e.g.,
\citealt{martinez07,dma08,donley08,hickox09,juneau11,trump13,mendez13}). For
example, due to obscuration bias, optical surveys miss $\approx$50\%
of the AGN population (e.g., \citealt{GA09,goulding10}), and even
X-ray surveys, which are relatively unaffected by moderate-levels of
obscuration, still miss a significant fraction ($\sim$20--50\%) of AGN
(e.g.,
\citealt{donley05,guainazzi05,park10,dma11,georgantopoulos13,wilkes13}).
To mitigate these biases, we harness the extensive multi-wavelength
datasets ({\it Spitzer}-IRAC; {\it WISE}; VLA-FIRST; {\it
  Chandra}-ACIS) falling in the SDSS, Bo\"{o}tes and DEEP2 footprints.

The proceeding sections describes our methodology (similar to that of
Hickox et al. 2009), for constructing a relatively large,
obscuration-independent, unbiased AGN sample in the three survey
regions (see also Fig.~\ref{fig:flowdiag}). For radio and X-ray AGN
identification we used a set of luminosity thresholds that are
sufficiently high to remove luminous star-forming systems, while at
infrared wavelengths we employed a well-established
AGN--star-formation ratio diagnostic to identify less luminous (or
obscured) AGN that are contributing significantly to the bolometric
luminosity of their host galaxies. Furthermore, for a flux-lmited
sample, there is tendency to identify AGN in more massive galaxies
(e.g., \citealt{aird12}). Hence, with the combination of the AGN
selection methods/wavelength ranges used here, we can mitigate many
potential baises and produce an almost complete and unbiased sample of
AGN spread across a wide dynamic range in accretion rates, obscuration
levels and AGN luminosities.

\subsection{Infrared AGN}

Deep mid-IR surveys performed with the WISE and {\it Spitzer}
observatories provide an almost obscuration-independent view of the
Universe and are now capable of resolving the bulk of the IR
background at $\sim$3--70$\mu m$ (e.g.,
\citealt{dole06,frayer06,ashby13}). These pointed, wide-field, and all-sky
surveys reveal large populations of dust-obscured starburst galaxies
and AGN. By contrast to standard AGN detection methods (e.g., at
radio; X-ray energies), which require a particular luminosity
threshold that is unlikely to be produced by non-accretion processes,
widely-used IR AGN identification diagnostics are predicated on the
contribution of the dust reradiated (intrinsic) AGN and star-formation
emission to the bolometric luminosity of a particular galaxy. Such a
diagnostic can often lead to the identification of more heavily
dust-obscured AGN than would be found at other wavelengths as both the
AGN and the star-formation can be extinguished by the dust located in
the circumnuclear region as well as the extended host galaxy.

\subsubsection{WISE-selected AGN in the SDSS}

The SDSS is covered in its entirety by the WISE mid-IR survey,
providing 4-band cryogenic photometry at 3.4, 4.6, 12 and 22$\mu
m$. We associated the SDSS sources with WISE counterparts using a
similar procedure to the one presented by \cite{dabrusco13}, which we
briefly recall here. For each SDSS source we searched for IR
counterparts in the WISE all-sky archive within circular regions of
variable radius \(R\) in the range between 0$''$ and 4$''$ with
increment \(\Delta R = 0.1''\).  For each radial distance, we
estimated the number of total (\(N_t(R)\)) and random (\(N_r(R)\))
matches together with the chance probability for the spurious
associations \(P(R)\). The random matches \(N_r(R)\) correspond to
those found by radially shifting the search region centroid by a
randomized distance between 10$''$ and 20$''$. The chance probability
for spurious associations \(P(R)\) was calculated as the ratio between
\(N_r(R)\) and the total number of sources. We calculated the
differences between the number of total matches at a given radius
\(R\) and those at \(R-\Delta R\) defined as \(\Delta N_t (R) =
N_t(R)-N_t(R-\Delta R)\), and the corresponding variation of the
random associations, \(\Delta N_r (R) = N_r(R)-N_r(R-\Delta R)\). At
radii larger than 3.1$''$, the increase in number of WISE sources that
were positionally associated with a SDSS source was systematically
lower than the increase in number of random associations. Thus,
3.1$''$ was chosen as the radial threshold for searching for
counterparts of SDSS sources in the WISE all-sky release. We
identified 43,171 single WISE counterparts matches within 3.1$''$ of
an SDSS galaxy, 445 multiple matches and 43,616 total matches, with a
chance probability for spurious associations of \(P(R<3.1'') \sim
0.5\%\). For the 445 ($\sim 1$\% of the total) sources with multiple
matches, we chose the counterpart with the smallest separation. The
extra $\sim 0.5$\% of multiple source matches, over the expected $\sim
0.5$\% spurious matches is due to the non-random distribution of
galaxies across the sky, i.e., galaxy clustering.

The majority of low-redshift extragalactic sources detected by WISE in
the 3.4, 4.6, and 12$\mu m$ bands are dominated at mid-IR wavelengths
by host-galaxy starlight. Because of this, a significant proportion of
optical emission-line classified AGN in the SDSS cannot be separated
from their star-forming and Low-Ionization Narrow Emission-Line Region
(LINER) galaxy counterparts using the typical WISE infrared
[3.4]--[4.6] vs [4.6]--[12] color-color diagram
\citep{yan12}. However, we may still adopt the relatively conservative
AGN selection criteria presented by \cite{mateos12} to select a mid-IR
AGN sample that is free from contamination by H{\sc ii} galaxies and
LINERs using the main SDSS matched galaxy sample, shown in
Fig.~\ref{fig:IR_AGN}a. Of the 311,635 SDSS galaxies in our matched
galaxy sample, 1,121 are selected as AGN using the \cite{mateos12}
WISE selection method. Fig.~\ref{fig:IR_AGN}a further confirms that
the mid-IR AGN selection criteria is not subject to contamination from
non-AGN systems. We predicted the color--color redshift tracks for a
set of starburst, star forming, spiral, and elliptical galaxies (see
\citealt{donley12}; hereafter, D12). These tracks were calculated by
convolving a set of infrared spectral energy distributions for the
archetypal template galaxies (J. Donley priv. conv.) with the WISE
photometric filter responses to predict their WISE colors in the
redshift range $z \sim 0.05$--0.2. None of these `non AGN' template
galaxies are expected to lie within the WISE AGN selection region in
the redshift range of the SDSS sample. Hence, our SDSS--WISE AGN
sample can be considered a clean representation of the overall
low-redshift IR AGN population.

\subsubsection{Spitzer-IRAC selected AGN in \bootes \& DEEP2}

Both the \bootes and DEEP2 surveys have dedicated infrared imaging
campaigns performed using the {\it Spitzer} Infrared Array Camera
(IRAC; \citealt{fazio04}) at 3.6, 4.5, 5.8 and 8.0$\mu m$. The \bootes
field is covered by the 8.5~deg$^2$ IRAC Shallow Survey
\citep{eisenhardt04} and the 10~deg$^2$ Spitzer Deep Wide-Field Survey
(SDWFS; \citealt{ashby09}). The final SDWFS catalog includes
$\approx$680,000 unique sources. The optical spectroscopic survey,
AGES \citep{kochanek12}, covers $\sim 7.3$~deg$^2$ of the SDWFS
field. The spatially associated optical--IR source catalog exists as
part of the AGES catalog, which we use here to provide the necessary
IRAC photometry for AGN selection for our matched galaxy catalog in
the \bootes field.

DEEP2 Fields 1, 2, and 4 have relatively complete four band coverage
with IRAC.\footnote{DEEP2 Field 3 has no current Spitzer IR
  photometry. For the purposes of homogeneity, as well as decreased
  sensitivity for distant objects, we do not include the available
  WISE photometry in DEEP2 Field 3 as this will potentially bias
  ensuing results to only the brightest systems, which may not be
  representative of the overall AGN/galaxy population.}  Spitzer IRAC
Guaranteed Time Observations of DEEP2 Field 1 were performed as part
of the Extended Groth Strip (EGS) campaign. The IRAC--EGS observations
cover a 10$' \times 120'$ contiguous region across the center of the
EGS survey and contain $\sim 57,500$ unique sources
\citep{barmby08}. In addition, General Observing IRAC data were
obtained for Fields 2 and 4 as part of Program 40689 PI G. Fazio and
Program 50660 PI C. Jones. These four band IRAC observations cover
0.92 deg$^2$ and 0.88 deg$^2$, respectively.

\begin{figure*}[t]
\includegraphics[width=\textwidth]{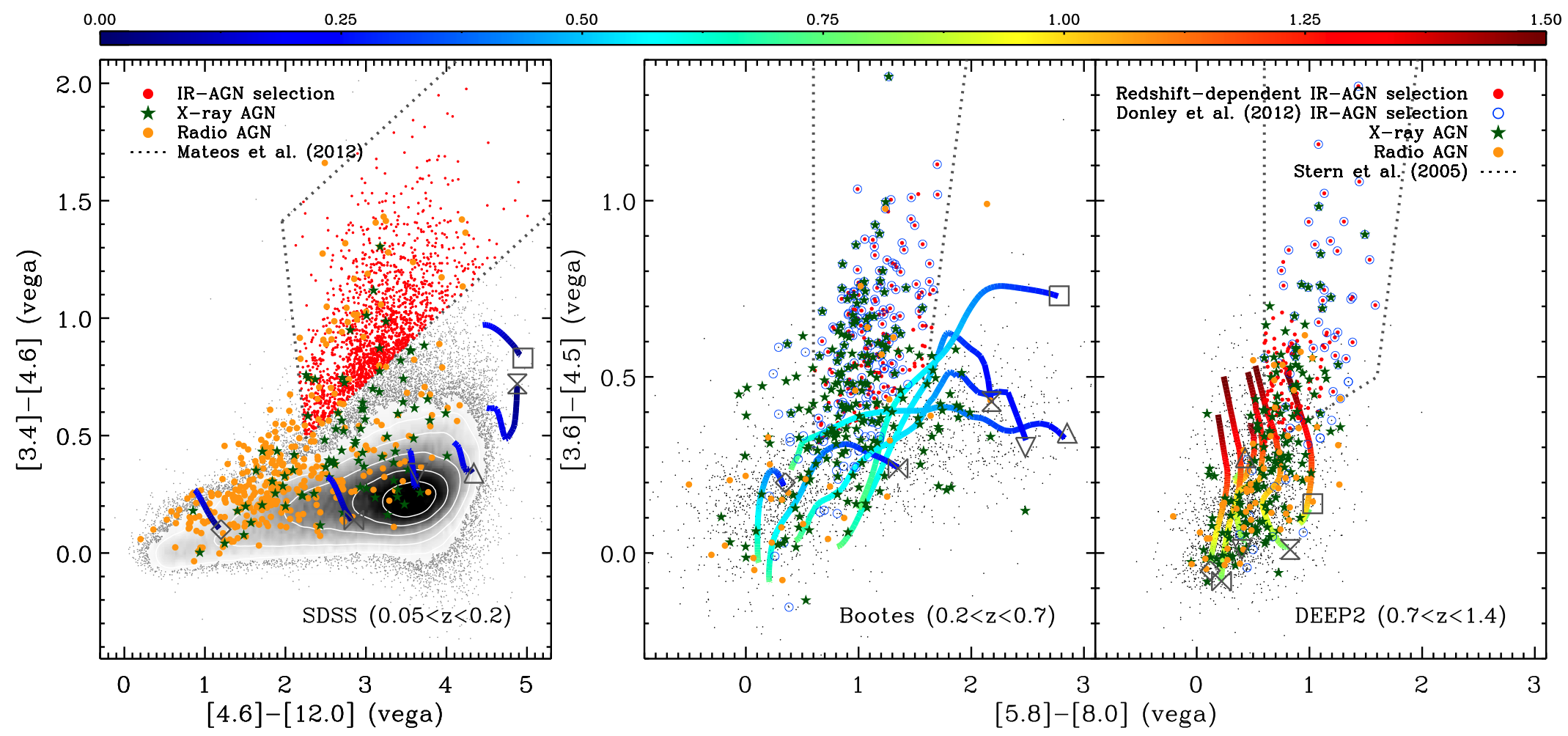}
\caption{{\it Common:} Redshift dependent spectral tracks for
  starburst (square, hourglass), spiral (triangle, upturned triangle,
  bow-tie) and elliptical (diamond) galaxies are shown. Radio and
  X-ray detected AGN are shown with orange circles and green stars,
  respectively. Infrared color--color diagnostic selected AGN are
  shown with red dots. Our choice of redshift dependent AGN selection
  regions (see main text) provides a clean AGN sample that does not
  included star-forming or passive galaxy (non AGN) interlopers. {\it
    a; left:} WISE infrared [3.4]--[4.6] versus [4.6]--[12.0] Vega
  magnitude color--color selection of galaxies and AGN detected in the
  SDSS for sources at $0.05<z<0.2$ (contours and gray dots). The AGN
  (red circles) lying within the AGN diagnostic selection region of
  Mateos et al. (2012) is shown with dotted lines. {\it b; center:}
  Spitzer IRAC infrared [3.6]--[4.5] versus [5.8]--[8.0] Vega
  magnitude color--color selection of galaxies selected in the \bootes
  survey in the redshift range, $0.2<z<0.7$ (gray dots). Candidate AGN
  identified using the Stern et al. (2005; dotted region) and/or
  Donley et al. (2012; blue open circles) infrared AGN color selection
  methods are additionally shown. Infrared selected AGN, after removal
  of star-forming interlopers are shown with red filled circles. {\it
    c; right:} Same as center panel except source sample is derived
  from the DEEP2 survey at $0.7<z<1.4$.}
\label{fig:IR_AGN}
\end{figure*}

For consistency between the surveys, the Spitzer IRAC data were
reduced following the method employed for the \bootes SDWFS
observations \citep{ashby09}, which we briefly recall here. Basic
Calibrated Data (BCD) images for the 3.6, 5.8 and 8.0$\mu m$ {\it
  Spitzer} IRAC observations were first post-processed using the MOPEX
software package\footnote{MOPEX is available from the Spitzer Science
  Center website at
  \url{http://irsa.ipac.caltech.edu/data/SPITZER/}.}, which removes
data artifacts such as stray-light, muxbleed, column pulldown, and
banding and re-calibrates the astrometry. The individual exposures
were then object masked and median stacked. These stacked images were
subtracted from the individual BCDs in each Astronomical Observation
Request (AOR) to remove long-term residuals caused by previous
observations of bright sources. This was not carried out for the
4.5$\mu m$ observations as the cryogenic 4.5$\mu m$ array does not
suffer from residual effects. The individual image tiles were combined
into full field mosaicked images, one for each band, using the Montage
toolkit. After the masking of the bright stars present in the
mosaicked images, sources were detected and extracted for each
individual wavelength band using the standard tool, SExtractor (v2.50;
\citealt{bertin96}). To allow a closer comparison with the sources
identified in the SDWFS data, we used SExtractor parameter settings
identical to those presented in Table 2 of \cite{ashby09}. This
detection method was verified by generating background and object
``check images''. Background subtracted photometry measurements were
extracted for both fields using 4$''$ diameter circular apertures and
the appropriate corrections applied as provided in the Spitzer Science
Center IRAC Instrument Handbook to scale the photometry to 24$''$
apertures (to match the methodology of Ashby et al. 2009). There are
38,851 and 34,228 unique sources detected in the [3.6] and [4.5] IRAC
bands in DEEP2 Fields 2 and 4, respectively.

Within the SDWFS--\bootes catalog, IRAC sources have optical
counterparts that were identified within a radial distance of $\delta
< 2.5''$, equivalent to a spatial distance of $\sim 12.5$ kiloparsecs
at the median redshift ($z\sim0.35$) of the \bootes sample. To
maintain consistency we spatially associated the IRAC sources in the
DEEP2 regions to the DEEP2 photometric catalog \citep{coil04} using a
matching radius of $\delta = 1.5''$ (equivalent to a spatial distance
of $\sim 12$~kpc at $z=1$). Of the IRAC sources detected with signal
to noise $\geq 5$ in all four IRAC bands, 3932, 1068, and 837 sources
were matched to an optical spectroscopic counterpart in the DEEP2
catalog in Fields 1, 2 and 4, respectively. We tested our choice of
matching radius using the IRAC--EGS data (Field 1), as the mid-IR data
within this field are the most sensitive and the optical photometry is
the most complete of the DEEP2 fields. We artificially shifted the
positions of the IRAC sources by 1$'$ to create a pseudo-random source
catalog and re-performed the counterpart matching. We identify 105
fake matches, suggesting a worst-case spurious matching fraction of
$\sim 2.7$\% for the IRAC counterparts. The spurious fraction in
Fields 2 and 4 are likely to be considerably smaller due to the
shallow IRAC data and pre-selection of $z>0.7$ targets in the optical
DEEP2 spectroscopic catalog.
 
We identified mid-IR AGN in \bootes and DEEP2 based on their position
in the typical IRAC color--color magnitude diagrams. AGN may be
identified in IRAC color space when the AGN-produced power-law can be
detected above the expected host-galaxy star-formation emission (i.e.,
a sufficiently high AGN to star-formation ratio). The two most widely
used formalisms are the AGN selection wedges presented by
\cite{lacy04} in $S_{8.0}/S_{4.5}$ versus $S_{5.8}/S_{3.6}$
color-space and \cite{stern05} in [3.6]--[4.5] versus [5.8]--[8.0]
color-space. These AGN selection criteria were most recently updated
by D12. Fig.~\ref{fig:IR_AGN}b,c presents the [3.6]--[4.5] versus
[5.8]--[8.0] distributions (in Vega magnitudes) for all IRAC sources
with \bootes and DEEP2 counterparts in our optical parent galaxy
catalog (see Section 2 and Appendix A).

AGN selection in IRAC color--color space provides a relatively clean
sample of AGN that is redshift independent. However, as shown by D12,
for specific redshift regimes and inherent dust extinctions,
contamination from starbursting or passive systems can be significant,
particularly for the Lacy criterion. In \bootes and DEEP2, we have the
advantage that the IRAC sources have well constrained spectroscopic
redshifts, and we may flag and remove (contaminant) sources that would
be otherwise identified as AGN candidates but lie in the region of
IRAC color parameter space that may be occupied by non AGN at a given
redshift.

Following D12, we track the expected positions of non AGN galaxies in
Stern color space using the spectral energy distributions (SEDs) of
the template galaxies described in the previous section. We convolved
the template SEDs with the IRAC filters to predict the position of
these non-AGN systems in the \bootes and DEEP2 redshift ranges. As we
show in Figs.~\ref{fig:IR_AGN}b,c, and has been shown previously by
D12 (see also \citealt{mendez13}), the lower portion of the Stern AGN
wedge suffers from significant contamination from pure starburst
galaxies at $z \sim 0.5$--0.7 and $z \gtrsim 1.2$. To mitigate the
contamination from starbursts to an IR AGN sample, we set a redshift
dependent threshold to the lower [3.6]--[4.5] cut that bounds the
Stern AGN wedge in the \bootes and DEEP2 IR samples, as follows:
{\footnotesize 
\begin{eqnarray}
&{\rm at~}& 0.4 < z \leq 0.6: [3.6]-[4.5] > 0.2([5.8]-[8.0]) + 0.28 \\
&{\rm at~}& 0.6 < z \leq 1: [3.6]-[4.5] > 0.2([5.8]-[8.0]) + 0.18 \\
&{\rm at~}& z > 1: [3.6]-[4.5] > 0.2([5.8]-[8.0]) + (0.5z - 0.32)
\end{eqnarray}
}
\noindent i.e., in the redshift range $0.6 < z < 1$ a standard Stern
AGN wedge may be used, as we predict it remains relatively free of
contamination from star-forming systems in this redshift
regime. Otherwise, in the redshift ranges where contamination is
expected to the AGN sample, we flag and remove these possible star
forming interlopers. Hence, starburst systems should not enter into
AGN selected catalogs. Thus, we select IRAC AGN in \bootes and DEEP2
using both our modified, redshift-dependent \cite{stern05} wedge in
[3.6]--[4.5] versus [5.8]--[8.0] color-space or the strict (redshift
independent) D12 criteria in $S_{8.0}/S_{4.5}$ versus
$S_{5.8}/S_{3.6}$ color-space, i.e., we consider a galaxy to contain
an IR AGN if it appears in either of the AGN wedges. Our final clean
IR AGN samples contain 151 and 135 sources in \bootes and DEEP2,
respectively.

\begin{figure*}[t]
\centering
\includegraphics[width=0.85\textwidth]{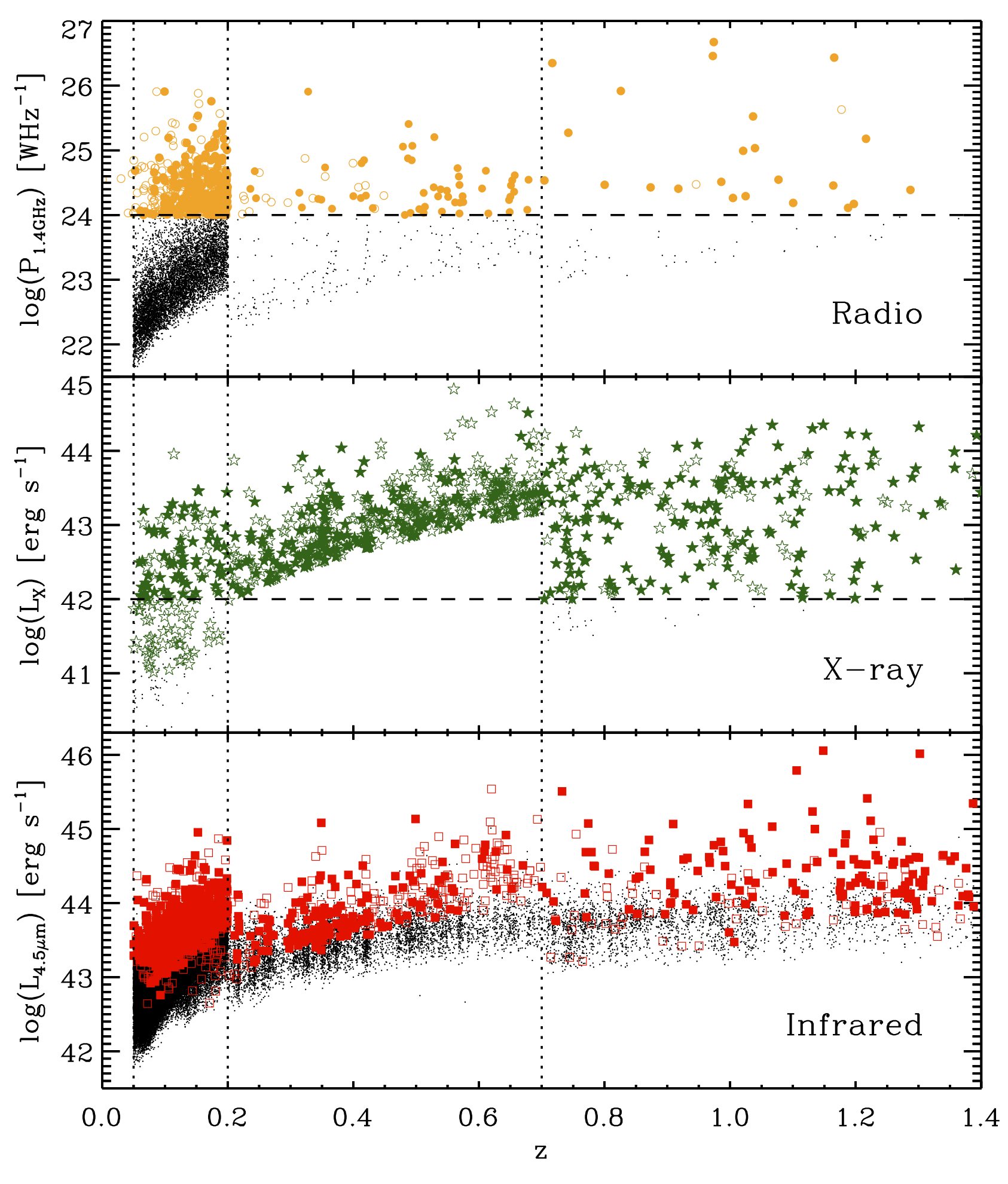}
\caption{Luminosity distributions of galaxies (dots) and radio, X-ray
  and IR AGN (filled circle; star; squares, respectively) in the
  survey samples. Radio AGN are selected at the threshold of $P_{\rm
    1.4Ghz} > 10^{24}$ W~Hz$^{-1}$, and X-ray AGN with $L_{\rm X} >
  10^{42} \ergps$. As part of the galaxy-matching procedure described
  in Section~2 and the Appendix, we remove galaxies that would
  hitherto not be detected in one of more of the three survey
  fields. For comparison purposes, those AGN hosts that are not
  included in the final optical-property matched sample are shown here
  with open symbols.}
\label{fig:z_lums}
\end{figure*}

\subsection{X-ray AGN}

Blank-field X-ray surveys arguably provide the most efficient and
unbiased AGN selection as X-ray emission is relatively unaffected by
moderate levels of obscuration and remains uncontaminated by
host-galaxy emission for luminosities in excess of $L_{\rm X,0.5-7keV}
\gtrsim 10^{42} \ergps$ (e.g., \citealt{moran99,lira02}). Here we
harness the unprecedented angular resolution provided by the {\it
  Chandra} X-ray Observatory to identify X-ray luminous AGN lying
within the SDSS, Bo\"{o}tes, and DEEP2 survey regions.

The Chandra Source Catalog (CSC; \citealt{evans10}) provides a uniform
catalog of high-significance X-ray sources for all observations
performed by {\it Chandra}. As part of the CSC analysis, a
cross-matched catalog is provided between the SDSS-DR7 release and the
current public release of the CSC (v1.1) using a Bayesian matching
routine and contains $\sim 8,950$ extragalactic sources.\footnote{For
  further information on the CSC--SDSS cross matched catalog see
  \url{http://cxc.harvard.edu/csc1/CSC-SDSSxmatch/}.} Despite covering
$\sim 130$~deg$^2$, the CSC--SDSS catalog cannot be considered a true
blank-field survey as the majority of observations within the SDSS
regions are individual pointed targets pre-selected for {\it Chandra}
observations. To guard against these potentially biased {\it Chandra}
targets, we flaged and removed all X-ray sources within 10$''$ of the
{\it Chandra} aim-point. Arguably, the choice of 10$''$ is
arbitrary. However, 10$''$ ensures that we remove X-ray sources that
are within the expected extent of a typical galaxy at $z\sim0.05$ (the
spatial scale at $z=0.05$ is $\sim$960pc/$''$) where we may be
sensitive to contamination from stellar-mass X-ray
binaries. Furthermore, our choice of 10$''$ allows for marginal
offcenter pointings of {\it Chandra} and discrepancies in perceived
target source position. We used the SDSS source-ID numbers to directly
match the remaining sources in the CSC--SDSS catalog to our matched
galaxy sample and found 242 optical--X-ray sources in common when
considering only those CSC--SDSS sources with a Bayesian matching
probability of $P > 0.9$ in the redshift range $0.05 < z < 0.2$.

\bootes and DEEP2 have almost complete coverage by {\it Chandra}, and
dedicated source matched catalogs for these surveys are available
\citep{murray05,kenter05,nandra05,laird09,kochanek12,goulding12b}. Here
we used the most recent cross-matched catalogs for \bootes and DEEP2
from \cite{kochanek12} and \cite{goulding12b}, respectively. The
X-\bootes and X-DEEP2 surveys are contiguous with $\sim 7.6$~deg$^2$
and $\sim 2.7$~deg$^2$ of their respective optical spectroscopic
surveys with total effective X-ray exposures in the range
5ks--1.1~Ms. When accounting for our source matching and selection
process, defined in Appendix A, there are 199 and 214 galaxies with
X-ray counterparts in the \bootes and DEEP2 fields, respectively.

For all of the X-ray sources in our matched catalogs, we applied X-ray
$K$-corrections to produce rest-frame X-ray luminosities,
\begin{equation} L_{\rm X,rest}({\rm 0.5-7}) = 4\pi d^2_l (1+z)^{\Gamma-2} S_{\rm
  X,obs}({\rm 0.5-7})
\end{equation}
\noindent where $d_l$ is the luminosity distance and assuming a
typical X-ray spectral slope of $\Gamma \sim 1.7$ \citep{tozzi06}. To
remove possible contamination from galaxies whose X-ray emission may
be dominated by high-mass X-ray binary systems and not AGN activity,
we selected only sources with $L_{\rm X,rest} \gtrsim 10^{42}
\ergps$. Our final SDSS, Bo\"{o}tes, and DEEP2 catalogs contain 73,
196, and 200 X-ray luminous AGN, respectively.

\subsection{Radio AGN}

Finally, we selected a sample of radio-luminous AGN within the three
surveys. As has been shown previously, radio-loud AGN are generally
identified in massive passive galaxies (e.g.,
\citealt{best05b,kauffmann08,hickox09}). Hence, and by contrast to IR
and X-ray AGN, radio-loud AGN may represent a separate AGN population
and/or a different evolutionary phase of galaxy and AGN activity.

As part of the current release of the FIRST radio catalog, a matched
SDSS (and 2MASS) source list is available directly from the SDSS
SkyServer. For the radio objects identified to be double or triple
component sources in the FIRST catalogs, we used the centroided
positions. For the lowest-significance FIRST sources (the typical
detection threshold is $\sim$0.75--1 mJy), the radio centroided
positions have typical random uncertainties of $<0.5''$, and
systematic uncertainties of $<0.05''$. We limited our selection to
those FIRST sources within 0.6$''$ of an SDSS counterpart to provide a
clean selection of SDSS--FIRST sources. If the FIRST radio positions
are systematically offset from the optical nucleus, projection effects
may adversely bias our cross-matching to only selecting counterparts
at the upper-end of our redshift range. Hence, we tested the use of
$\delta = 0.6''$ as compared to a larger radius of $\delta = 3.0''$,
while we found a significant increase in the number of possible
counterparts for the larger radius, the overall redshift distribution
of the counterparts remained unchanged. Thus our choice of matching
radius can be considered clean and robust.

For the $\sim$6\% of FIRST sources with multiple possible SDSS
counterparts in the SDSS--FIRST SkyServer catalog, we chose the SDSS
galaxy with the smallest separation from the FIRST source. Of these
FIRST sources with multiple declared SDSS matches, each has one
counterpart within $<0.6''$, and other (spurious) SDSS counterparts at
1.5--8$''$. There are 6,959 FIRST sources with SDSS counterparts
within $\delta = 0.6''$ in the redshift range, $0.05<z<0.2$.

The Deep Westerbork 1.4GHz radio telescope (WSRT) catalog of the
\bootes field \citep{devries02} contains 3172 $5 \sigma$ detected
unique sources down to a $1\sigma$ background limiting flux of
$28\mu$Jy. We matched all radio sources to optical counterparts within
1.5$''$ which accounts for the larger systematic positional
uncertainty in the Westerbork imaging (median $\sim 0.5''$) compared
to the $0.05''$ systematic uncertainty in the VLA FIRST data in our
SDSS matched catalog. There are 259 WSRT radio sources with
spectroscopic \bootes counterparts at $0.2<z<0.7$.

Finally, the DEEP2 fields are covered by the FIRST VLA survey, while a
dedicated deep (50$\mu$Jy) 20cm VLA survey, AEGIS20
\citep{ivison07,willner12} additionally covers the EGS (DEEP2
Field~1). Following our strategy for cross-matching and selection of
sources between VLA and SDSS, we selected all centroided radio sources
within 0.6$''$ of a DEEP2 optical counterpart. We found 180 radio
sources in our cross-matched VLA--DEEP2 catalog at $0.7<z<1.4$.

Using our radio--optical cross-matched catalogs, we perform the
necessary rest-frame radio $K$-corrections and calculate radio
luminosities,

\begin{equation}
L_{\nu}(\nu_{\rm rest}) = 4\pi d^2_l (1+z)^{\alpha-1} S_{\nu}(\nu_{\rm obs})
\end{equation}

\noindent where $S_{\nu}$ is the FIRST radio flux at 1.4GHz and
assuming a radio spectral slope of $\alpha = 0.5$. To guard against
the selection of luminous starburst systems, we selected as radio AGN
only sources with $P_{\rm 1.4GHz} > 10^{24}$W~Hz$^{-1}$. Assuming a
typical radio to star-formation rate conversion of log(SFR)$={\rm
  log}(\nu L_{\rm 1.4GHz})-37.07$ \citep{condon92}, the required
star-formation rate for a starburst galaxy to erroneously meet our AGN
selection threshold is $SFR \sim 1200 M_{\odot}{\rm yr}^{-1}$, typical
of a Hyper Luminous Infrared Galaxy. In the redshift ranges considered
here, the space density of star-forming galaxies with $L_{\rm 1.4GHz}
> 10^{24}$W~Hz$^{-1}$ is predicted to be $\Phi \sim 6 \times 10^{-7}$
at 0.05$< z <$0.2, $\sim 7 \times 10^{-6}$ at 0.2$< z < $0.7, and
$\sim 10^{-5}$~Mpc$^{-3}$~dex$^{-1}$ at 0.7$< z <$1.4
\citep{smolcic09}. Thus, we estimate that $\sim 5$, $\sim 2$ and $\sim
3$ of the $P_{\rm 1.4GHz} > 10^{24}$W~Hz$^{-1}$ radio sources
identified in the SDSS, Bo\"{o}tes, and DEEP2, respectively, will have
their radio emission dominated by rapid star-formation and not AGN
activity. Hence, we may consider $P_{\rm 1.4GHz} > 10^{24}$W~Hz$^{-1}$
a conservative and robust choice of luminosity threshold to identify
AGN. Furthermore, the relatively modest-depth of the FIRST survey
($P_{\rm limit}(z\sim 1) \approx 2 \times 10^{24}$W~Hz$^{-1}$) does
not affect our ability to detect luminous radio-loud AGN in the DEEP2
survey as the majority ($\sim$70\%) of the radio sources identified in
the DEEP2 fields have rest-frame luminosities of $P_{\rm 1.4GHz} \sim
(0.3$--$4) \times 10^{24}$W~Hz$^{-1}$.

Our choice of radio luminosity threshold will select against
moderately radio-quiet AGN, of which there are a substantial number in
the SDSS ($\sim 2000$; see \citealt{mullaney13}). However, our goal
here is not to select a highly complete sample of radio AGN but
instead one that is free from contamination from star-formation
dominated galaxies so that we may accurately compare the AGN host
galaxies and their subsequent evolution. Therefore, our final
conclusions towards the evolution of radio AGN are restricted to only
the highest luminosity population.

The redshift distributions for the IR, X-ray, and radio AGN samples
are presented in the lower panels of Fig.~\ref{fig:z_hists} and the
redshift--luminosity distributions are shown in
Fig.~\ref{fig:z_lums}. In general, we span similar AGN luminosity
regimes throughout the three surveys. Despite the differing
sensitivities of the IRAC observations in DEEP2 Field 1 and Fields 2
and 4, it is important to consider these observations together as a
combined sample because the limited survey area of Field 1 does not
yield sufficient sources to cover the full luminosity range found in
\bootes and the SDSS. Finally, due to the sensitivity limit of the
{\it Chandra} observations in \bootes, we are incomplete for AGN with
$L_{\rm X} \sim 10^{42}$--$10^{43} \ergps$ at $0.4<z<0.7$. In this
redshift regime we are restricted towards relatively modest to high
Eddington ratio sources in low mass galaxies, due to our ability to
detect only the most luminous AGN with smaller black holes. However,
this does not limit our conclusions towards the identification of the
effects of AGN feedback, as it is high Eddington ratio AGN that are
expected to being undergoing significant evolutionary events.

\begin{figure*}[t]
\includegraphics[width=\textwidth]{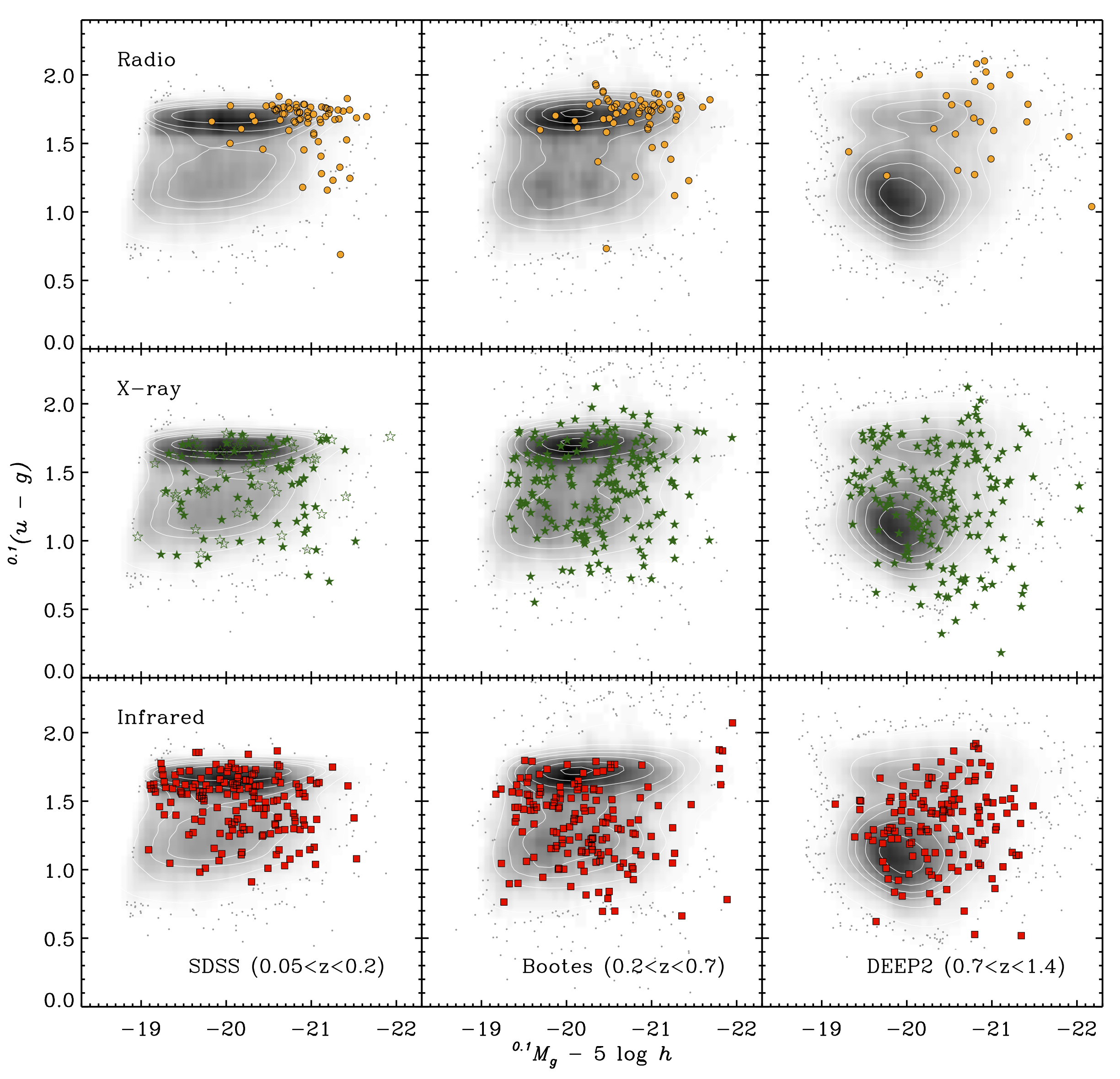}
\caption{Rest frame optical color--magnitude distributions of AGN
  hosting galaxies at $z<0.2$ (SDSS; left column), $0.2<z<0.7$
  (Bo\"{o}tes; center column) and $0.7<z<1.4$ (DEEP2; right
  column). Photometry for the galaxy samples have been $K$-corrected
  and converted to equivalent SDSS $u$ and $g$ filters at $z=0.1$, and
  absolute $g$-band magnitudes are corrected for passive evolution
  (e.g., Blanton et al. 2003; 2007). The color--magnitude diagrams for
  the selection-matched galaxy populations, in each redshift range,
  are shown with grayscale contours. Overlaid on the optical
  color--magnitude diagrams are radio ($P_{\rm 1.4GHz} \gtrsim
  10^{24}$W~Hz$^{-1}$), X-ray ($L_X \gtrsim 10^{42} \ergps$) and
  IR-selected AGN which are shown with filled orange dots, green stars
  and red squares, respectively. Low-redshift ($z<0.2$) X-ray AGN
  identified in the SDSS (with $L_X \sim 10^{41}$--$10^{42} \ergps$)
  are shown with open green stars. From Markov-Chain Monte Carlo
  2-dimensional Kolmogorov-Smirnov testing, we find that in each
  redshift regime, radio AGN are almost exclusively hosted in passive
  red galaxies with little evidence for evolution in their host-galaxy
  color--magnitude distributions. Furthermore, we find that at all
  redshifts X-ray AGN are spread throughout the color--magnitude
  diagram, with similar distributions at all redshifts, while IR AGN
  may show marginal evidence for evolution between $0.7<z<1.4$ and
  $0.2<z<0.7$.}
\label{fig:col_mag}
\end{figure*}

\section{The Evolution of AGN Host-Galaxies to {\footnotesize z}$<1.4$}
\label{sec:results}

Evidence is now emerging that long-term AGN activity traces star
formation in galaxies, suggesting a strong link between the properties
of AGN and their host galaxies (e.g., \citealt{chen13,hickox13} and
references therein). Links between AGN accretion type and the
evolution of their host are likely to be reflected in the host
characteristics. For example, based on sensitive multi-wavelength
observations in the nearby Universe, the most rapidly growing BHs are
preferentially found in relatively low mass, star-forming galaxies
(e.g., \citealt{GA09}) and are rarely found in large, passive systems
(e.g., \citealt{hickox09}), supporting a paradigm where-by a
substantial gas reservoir is required to simultaneously grow the BH
and sustain the birth of new stars. Furthermore, at more moderate
redshifts ($z \sim 1$--2), the most luminous, gas and dust-rich
starburst galaxies are commonly associated with powerful, often
heavily-obscured AGN (e.g., \citealt{Page04, dma05}).

\subsection{The color--magnitude distribution of AGN hosts to $z<1.4$}
\label{subsec:col_mag}

A particularly powerful diagnostic for understanding the most general
properties of galaxies (i.e., luminosity; mass; evolution) is their
distribution in rest-frame color and absolute magnitude parameter
space. The galaxy color-magnitude diagram is bimodal (e.g.,
\citealt{strateva01,hogg03,baldry04,faber07}) with a distinct
separation between passive red-sequence galaxies and the blue cloud of
actively star-forming systems.

To determine whether a relationship exists between galaxies that host
AGN and their galaxy properties, and more importantly, to assess if
such a relationship evolves with redshift, we can examine the
distribution of AGN in host galaxy color--magnitude
space. Fig.~\ref{fig:col_mag} presents the $^{0.1}(u - g)$ versus
$^{0.1}M_g$ color--magnitude distributions for the galaxies in our
property-matched optical catalog separated in three redshift slices,
$0.05<z<0.2$, $0.2<z<0.7$ and $0.7<z<1.4$. Overlaid in the diagram are
those sources detected as AGN in the radio, X-ray, and/or mid-IR
surveys.\footnote{For presentation purposes only, we show a random
  subset of the radio and IR AGN detected in the SDSS. However, for
  all of our cross-correlation analyses we utilize the entire AGN
  ensemble.}

As expected, given the strongly evolving galaxy luminosity function
from $z \sim 1.5$ to the present-day combined with the bias in the
original DEEP2 selection against red galaxies
(\citealt{willmer06,faber07}, and references there-in),
Fig.~\ref{fig:col_mag} shows that the galaxies identified in DEEP2 (at
$0.7<z<1.4$), are heavily dominated by blue-cloud systems. Conversely,
the SDSS galaxies (at $0.05<z<0.2$) are dominated by those in the red
sequence, while \bootes galaxies are undergoing a transition due to
the assembly of dense group and cluster environments, which host the
majority of red-sequence galaxies in the present-day
Universe. Previous analysis of the color--magnitude distributions of
SDSS and DEEP2 galaxies have found fundamentally similar results
(e.g., \citealt{blanton06}) in that galaxies appear intrinsically
bluer at higher redshift, and these blue galaxies compose the dominant
fraction of galaxies in DEEP2 \citep{willmer06}. By applying the
evolution corrections (see Appendix A) to the red sequence and blue
cloud galaxies of \cite{blanton06}, as well as the like-for-like
galaxy matching between SDSS, \bootes and DEEP2, we find similar
overall color--magnitude distributions to those found previously.

The previous study of AGN within the spectroscopically identified
galaxies at $0.25<z<0.8$ in \bootes \citep{hickox09} investigated the
average $^{0.1}(u - r)$ distributions of radio, X-ray, and IR
AGN. They found that radio AGN are generally confined to the most
massive red-sequence galaxies (see also
\citealt{dunlop03,best05b}). Furthermore, they found that while X-ray
AGN are spread across a wide range in host galaxy color (see also
\citealt{xue10}), the median color of an X-ray AGN is that of a
`green-valley' galaxy (i.e., a galaxy lying between the blue-cloud and
red-sequence; see also
\citealt{nandra07,schawinski07,coil09,georgakakis11}), and finally,
they found that IR AGN are generally blue-cloud galaxies and/or less
massive satellite galaxies that populate the faint end of the red
sequence (but see also \citealt{mendez13}). After imposing our strict
sample selection techniques to allow our comparison to SDSS and DEEP2,
we obtain almost identical results to \cite{hickox09} for the three
AGN detection techniques. We also confirm the large range in galaxy
color ($\sim 1.6$ magnitudes) for X-ray AGN, spanning almost the
entire color--magnitude diagram without evidence for a bimodal
distribution as observed in the main galaxy population. Although we
select only a subset of the main galaxy populations available in each
survey field, it appears that, to first-order, our sample is a
representative sub-group that accurately reproduces previous results
for both AGN (in \citealt{hickox09}) and the main SDSS--DEEP2 galaxy
population (in \citealt{blanton06}), serving as a good-sanity check
for our methodology.

Fig.~\ref{fig:col_mag} further shows that in all three redshift
ranges, the radio AGN are primarily confined to high luminosity (high
galaxy mass) red-sequence galaxies, while the IR and X-ray AGN are
distributed throughout the color--magnitude diagram. Qualitatively,
the X-ray AGN have a similar color--magnitude distribution across the
SDSS, \bootes and DEEP2 surveys, i.e., the median host-galaxy colors
of X-ray AGN appear similar at all redshifts (to $z\sim
1.4$). Furthermore, the color--magnitude distributions of radio AGN
appear the same at all redshifts. Such results may be in conflict with
the evolving galaxy population where the blue-cloud galaxies are
transitioning onto the red sequence throughout these epochs. By
contrast, there is qualitative evidence that the distributions of IR
AGN change between \bootes and DEEP2, with a further marginal change
in the IR AGN distributions between \bootes and the SDSS. 

An alternate line of investigation would be to trace the AGN host
galaxies as a function of stellar mass (e.g., X-ray AGN host galaxy
properties; \citealt{aird12}), although given the limited photometric
coverage of the surveys considered here, stellar mass estimates become
inherently degenerate and inaccurate in these sources. We cannot be
complete to a single stellar mass limit across the entire redshift
range in either red sequence or blue cloud galaxies because our galaxy
property corrections unphysically alter the implied stellar mass
function for these sources. However, given that our final luminosity
matched sample is designed in the SDSS rest-frame, we can use the
mass-to-light conversions of \cite{bell03} to at least tentatively
place our sample into the stellar-mass regime to allow a first-order
comparison to previous investigations. Based on our limiting galaxy
color and luminosities of $^{0.1}(u-g)\sim0.60$ and 1.65 and $M_g \sim
-18.9$ and -19.1 for red sequence and blue cloud galaxies,
respectively, we estimate that at $z<0.2$ our sample is equivalent to
a stellar-mass limited sample with $M_{*,blue} \approx 8 \times 10^9
M_{\odot}$ and $M_{*,red} \approx 2.6 \times 10^{10} M_{\odot}$.

\subsection{Evidence for AGN host galaxy evolution}
\label{subsec:hostgal_evol}

We performed a two-dimensional Kolmogorov-Smirnov test
\citep{peacock83} between AGN populations at given epochs to
statistically analyze their distribution in color--magnitude space and
search for statistical similarities. Given the varying degree of
uncertainty towards the individual optical $K$-corrections between the
three surveys, as well as the relatively small number of sources for a
given AGN type and redshift bin, we employed a 2-D K-S test within a
Markov-Chain Monte-Carlo routine combined with a bootstrap analysis to
robustly calculate the statistical association significance between
the AGN samples.

\subsubsection{Constructing a robust Two-dimensional
  Kolmogorov-Smirnov test of AGN host distributions}

For individual sources, we modeled the uncertainties in the
color--magnitude measurements as two-dimensional Gaussians with
full-width half-maxima values of 2.35$\sigma_{\rm i,j}$, where
$\sigma_{i}$ and $\sigma_{j}$ are the uncertainties on the
$^{0.1}(u$--$g)$ color and absolute $g$-band magnitudes,
respectively. These uncertainties were derived directly from the
template fitting during the {\sc kcorrect} process described in the
Appendix. Typically, the uncertainties on the galaxy colors are $\sim
0.08$, $\sim 0.10$, $\sim 0.19$ for sources in the SDSS, \bootes and
DEEP2 surveys, respectively. For each AGN population, selected at a
given wavelength and redshift (i.e., each panel of
Fig.~\ref{fig:col_mag}), we constructed 10,000 random samples from the
individual source model uncertainties. We used a flat Heaviside prior
to limit the random sampling to the range in color--magnitude
measurements found in the main parent population distribution. To
guard against outliers and false detections, which can be a particular
problem during 2-D K-S testing, we used a heuristic bootstrap to
randomly select 90\% of the detected AGN in each panel and performed a
2-D K-S test between each of the panels for the 10,000 random samples
drawn from the individual measurement uncertainties. More
specifically, after each K-S test, we used a random-walk
Metropolis-Hastings sampler as part of the bootstrap routine to
re-evaluate and adaptively weight each individual source based on its
overall contribution to the K-S result. Using the 10,000 simulated
measurements, the M-H sampler has the effect of minimizing the
contribution of an outlier (i.e., a possible false optical
counterpart) to the K-S probability. We built full probability
distributions of the 9,000 (10,000 x 90\%) K-S test results for each
of the AGN populations in the given redshift slices. We then calculate
the maximum a posteriori (MAP) values for each set of AGN populations
directly from the peak of the probability distributions.\footnote{The
  MAP values are similar in principle to a maximum likelihood value
  with the added stipulation of our choice of Heaviside prior to limit
  the distribution of color--magnitude measurements.}

\begin{table*}
\footnotesize
\begin{center}
\setlength{\tabcolsep}{1mm}
\setlength{\fboxrule}{1pt}
\setlength{\fboxsep}{4pt}
\renewcommand{\arraystretch}{1.3}
\caption{Cross-Correlation Analysis of AGN host-galaxy color--magnitude distributions\label{tbl:2d_ks}}
\begin{tabular}{|lc|ccc|cccccc}
\cline{3-11}
\multicolumn{2}{c|}{} &
\multicolumn{3}{c|}{SDSS} &
\multicolumn{3}{c|}{\bootes} &
\multicolumn{3}{c|}{DEEP2} \\
\multicolumn{2}{c|}{} &
\multicolumn{3}{c|}{$0.05<z<0.2$} &
\multicolumn{3}{c|}{$0.2<z<0.7$} &
\multicolumn{3}{c|}{$0.7<z<1.4$} \\
\multicolumn{2}{c|}{} &
\multicolumn{1}{c}{IR} &
\multicolumn{1}{c}{X-ray} &
\multicolumn{1}{c|}{Radio} &
\multicolumn{1}{c}{IR} &
\multicolumn{1}{c}{X-ray} &
\multicolumn{1}{c|}{Radio} &
\multicolumn{1}{c}{IR} &
\multicolumn{1}{c}{X-ray} &
\multicolumn{1}{c|}{Radio} \\[1.2ex]
\tableline
 & & & & & \multicolumn{3}{c|}{} & \multicolumn{3}{c|}{}  \\
\multirow{7}{*}{DEEP2} & \multirow{2}{*}{Radio} & \multirow{2}{*}{\fcolorbox{red}{white}{$<0.07$}} & $4.1 \times 10^{-3}$ & \multirow{2}{*}{\fcolorbox{lime}{white}{0.29}} & \multirow{2}{*}{\fcolorbox{red}{white}{$<10^{-3}$}} & $3.0 \times 10^{-3}$ & \multicolumn{1}{c|}{\multirow{2}{*}{\fcolorbox{lime}{white}{0.36}}} & \multirow{2}{*}{\fcolorbox{red}{white}{$<0.07$}} & \multirow{2}{*}{\fcolorbox{red}{white}{$<0.04$}} & \multicolumn{1}{c|}{\multirow{2}{*}{-}} \\
                       &                        &  & ($<0.02$) &  &  & ($<0.02$) & \multicolumn{1}{c|}{} &  &  & \multicolumn{1}{c|}{} \\[1.75ex]
                       & \multirow{2}{*}{X-ray} & $1.8 \times 10^{-3}$ & \multirow{2}{*}{\fcolorbox{lime}{white}{0.25}} &  \multirow{2}{*}{\fcolorbox{red}{white}{$< 10^{-5}$}} & $0.04$ & \multirow{2}{*}{\fcolorbox{lime}{white}{0.12}} & \multicolumn{1}{c|}{ \multirow{2}{*}{\fcolorbox{red}{white}{$< 10^{-5}$}}} & \multirow{2}{*}{\fcolorbox{lime}{white}{0.20}} & \multirow{2}{*}{-} & \multicolumn{1}{c|}{} \\
                       &                        & ($<0.26$) &  &  & ($<0.70$) &  & \multicolumn{1}{c|}{} &  &  & \multicolumn{1}{c|}{} \\[1.75ex]
                       & \multirow{2}{*}{IR}    & $1.0 \times 10^{-3}$ & \multirow{2}{*}{\fcolorbox{lime}{white}{0.17}} & \multirow{2}{*}{\fcolorbox{red}{white}{$<2\times 10^{-5}$}} & $0.04$ & $0.07$ & \multicolumn{1}{c|}{\multirow{2}{*}{\fcolorbox{red}{white}{$<10^{-5}$}}} & \multirow{2}{*}{-} &  & \multicolumn{1}{c|}{} \\
                       &                        & ($<0.11$) &  &  & ($<0.78$) & ($<0.81$) & \multicolumn{1}{c|}{} &  &  & \multicolumn{1}{c|}{} \\[1.75ex]
\cline{1-11}
 & & & & & \multicolumn{3}{c|}{}   \\
\multirow{7}{*}{\bootes} & \multirow{2}{*}{Radio} & \multirow{2}{*}{\fcolorbox{red}{white}{$< 10^{-5}$}} &  \multirow{2}{*}{\fcolorbox{red}{white}{$<3\times 10^{-3}$}} & \multirow{2}{*}{\fcolorbox{lime}{white}{0.30}} & \multirow{2}{*}{\fcolorbox{red}{white}{$< 10^{-5}$}} & \multirow{2}{*}{\fcolorbox{red}{white}{$< 10^{-4}$}} & \multicolumn{1}{c|}{\multirow{2}{*}{-}} &  &  &  \\
                         &                        &  & &  &  &  & \multicolumn{1}{c|}{}                   &  &  &  \\[1.75ex]
                         & \multirow{2}{*}{X-ray} & $0.02$ & \multirow{2}{*}{\fcolorbox{lime}{white}{0.52}} & \multirow{2}{*}{\fcolorbox{red}{white}{$<0.03$}} & $0.02$ & \multirow{2}{*}{-} & \multicolumn{1}{c|}{} &  &  &  \\
                         &                        & ($<0.58$) &  &  & ($<0.50$) &                    & \multicolumn{1}{c|}{} &  &  &  \\[1.75ex]
                         & \multirow{2}{*}{IR}    & $0.01$ & \multirow{2}{*}{\fcolorbox{lime}{white}{0.10}} & \multirow{2}{*}{\fcolorbox{red}{white}{$<10^{-5}$}} & \multirow{2}{*}{-} &  & \multicolumn{1}{c|}{} &  &  &  \\
                         &                        & ($<0.34$) &  &  &                    &  & \multicolumn{1}{c|}{} &  &  &  \\[1.75ex]
\cline{1-8}
 & & & & &   \\
\multirow{7}{*}{SDSS} & \multirow{2}{*}{Radio} & \multirow{2}{*}{\fcolorbox{red}{white}{$<2 \times 10^{-3}$}} & \multirow{2}{*}{\fcolorbox{red}{white}{$<0.04$}} & \multirow{2}{*}{-} &  &  &  &  &  &  \\
                      &                        &  &  &                    &  &  &  &  &  &  \\[1.75ex]
                      & \multirow{2}{*}{X-ray} & \multirow{2}{*}{\fcolorbox{lime}{white}{0.11}} & \multirow{2}{*}{-} &  &  &  &  &  &  &  \\
                      &                        &  &                    &  &  &  &  &  &  &  \\[1.75ex]
                      & \multirow{2}{*}{IR}    & \multirow{2}{*}{-} &  &  &  &  &  &  &  &  \\
                      &                        &                    &  &  &  &  &  &  &  &  \\[1.75ex]
\cline{1-5}
\end{tabular}
\end{center}
{\bf Notes:-}
Maximum a posteriori (MAP) value produced from a Markov-Chain
Monte-Carlo two-dimensional Kolmogorov-Smirnov and bootstrap
analysis which calculates if two AGN populations at a
given epoch are drawn from the same parent AGN population. For
cross-associations with $KS({\rm MAP})<10^{-3}$, the $3\sigma$
upper-limit is given and was calculated from the KS value encompassed by
the lower 99.7\% of the 10,000 random samples in the 2-D KS test (red
box). $KS({\rm MAP}) \gtrsim 0.2$ indicates a
high-likelihood of a statistical correlation between the
color--magnitude distributions of the AGN samples at the 99\%
confidence level or higher (green box). For cross-associations with
weak/inconclusive statistical significances, we provide the $KS({\rm
  MAP})$ and $KS(3\sigma)$ (in parentheses).
\renewcommand{\arraystretch}{1}
\end{table*}

We tested our methodology by constructing mock samples of correlated
and non-correlated data. These mock samples were built by mixing
directly-correlated 2-D data with random `noise' in pre-defined
proportions while maintaining overall 2-D distributions that were
similar to the main bi-modal galaxy distribution readily observed in
the three surveys. The K-S distributions from our mock catalogs were
characterized by strongly skewed log-normal (almost one-sided)
distributions with extended tails towards low probabilities and rapid
declines from the peak of the K-S distributions (the MAP values)
towards marginally larger probabilities. Mock samples consisting of
100\% non-correlated data, the MAP value was invariably $KS({\rm MAP})
\ll 0.001$. For partially correlated populations (i.e., pre-determined
mixtures of correlated and non-correlated data) the MAP was in the
range $0.001 < KS({\rm MAP}) < 0.02$ for 67\% sample mixtures, $0.02 <
KS({\rm MAP}) < 0.2$ for 95\% mixtures, and $KS({\rm MAP}) > 0.20$ for
99.9\% mixtures. Hence, we can consider a positive association between
two populations at the $1 \sigma$ confidence level to have $0.001 <
KS({\rm MAP}) < 0.02$.

\subsubsection{Results of the 2-D K-S Evaluation}

Table~\ref{tbl:2d_ks} presents the MAP values of the K-S distributions
calculated for the cross-association of each panel in
Fig.~\ref{fig:col_mag} (i.e, we calculate the 2-D K-S MAP value for
AGN$_{\rm SDSS,radio}$ versus AGN$_{\rm Bootes,radio}$; AGN$_{\rm
  SDSS,radio}$ versus AGN$_{\rm DEEP2,radio}$; AGN$_{\rm
  Bootes,X-ray}$ versus AGN$_{\rm Bootes,IR}$ etc.). For
cross-associations which yield $KS({\rm MAP}) < 10^{-3}$ during our
2-D K-S analysis, we instead provide the $3\sigma$ upper-limit of the
KS value (i.e., the KS value encompassed by the lower 99.7\% of the
random draws). Furthermore, for those cross-associations that yield a
relatively ambiguous result (i.e., only partial/weak evidence for
similarities in their 2-D distributions in color--magnitude space) we
provide both the $KS({\rm MAP})$ and the $3\sigma$ upper-limit in
parentheses. For example, the cross-association of the color-magnitude
distributions observed for X-ray AGN identified in DEEP2 (at
$0.7<z<1.4$) compared with the distribution for IR AGN identified in
the SDSS (at $0.05 < z <0.2$) produces $KS({\rm MAP}) \sim 1.8 \times
10^{-3}$ and a $3\sigma$ upper-limit of $KS(3\sigma) < 0.26$.

The results presented in Table~\ref{tbl:2d_ks} confirm our earlier
impression that at any given redshift, the host galaxies of radio AGN
appear to be drawn from the same parent population, which maintains
very similar color--magnitude distributions with a statistical
significance of $P>95$\%. Concurrently, the host galaxies of X-ray AGN
also appear to maintain similar distributions across the redshifts,
and these are not correlated with the host galaxies of radio
AGN. Hence, we find strong statistical evidence that the
color--magnitude distributions of AGN host galaxies have not evolved
significantly in the last 9~Gyrs of Cosmic history. Furthermore, as
already noted by Hickox et al. (2009), the radio and IR AGN
populations appear anti-correlated in \bootes galaxies. From our
statistical analysis, a similar result appears to hold across all of
the redshift-regimes considered here: at any given redshift, IR and
radio-selected AGN exist in separate host-galaxy populations. Such a
result is consistent with the notion that the accretion processes are
distinctly different between the X-ray/IR AGN population and the
majority of the radio AGN population. As the evolution of galaxies is
expected to be a long term process, and the host galaxies of
radiatively efficient and inefficient AGN appear to be separate at a
given redshift, we find little evidence for a `flip-flop' between
accretion modes within the same galaxies. The original DEEP2 galaxy
rest-frame UV selection and spectroscopic selection results in a
marginal bias towards the selection of blue galaxies, particularly at
$z > 1.1$ (see Appendix A and \citealt{faber07}). Thus, to reconcile
the observed anti-correlation between IR and radio AGN, red galaxies
hosting IR AGN (and not radio-loud AGN) would need to be
systematically missed throughout the overall population. On the
otherhand, IR AGN generally present strong AGN emission lines at
optical wavelengths, which would provide a bias towards their
detection and inclusion in the sample. Hence, despite the deficit in
DEEP2 red galaxies, it appears that the observed anti-correlation
between the AGN populations is relatively robust.

The lack of a 2-D correlation between IR and radio sources may only
hold for low-excitation (LEX) radio AGN ($P_{\rm 1.4GHz} \sim
10^{24}$--$10^{25}$ W~Hz$^{-1}$). Those radio AGN in a high-excitation
mode (HEX; with $P_{\rm 1.4GHz} > 10^{25}$ W~Hz$^{-1}$) appear to be
confined to bluer galaxies ($u-g \sim 1.3$) than the LEX radio AGN
(e.g., \citealt{smolcic09b}) and have (qualitatively) similar
host-galaxy distributions to the X-ray and IR AGN. However, given the
low-number (13) of HEX radio AGN present in \bootes and DEEP2, we lack
the source statistics to place robust constraints on any correlation
on the evolution of HEX radio sources and X-ray/IR AGN.

DEEP2 reveals a substantial number of IR (35) and X-ray (39) AGN in
luminous red-sequence galaxies at $z \sim 1$ that are absent at $z <
0.7$. Analysis of the luminosity distributions of these particular
DEEP2 AGN finds that the galaxies are relatively high-luminosity
systems with $L_{\rm 4.5\mu m} > 5 \times 10^{44} \ergps$. Their
absence in SDSS and \bootes may be a luminosity downsizing effect as
only two (three) AGN with similar luminosities are present in the SDSS
(Bo\"{o}tes) samples. Conversely, there is a population of IR AGN in
low-luminosity ($M_g > -20$) red-sequence galaxies at $z<0.7$ that do
not appear at higher redshifts in DEEP2. However, we detect a small
coincident population of 18 low X-ray luminosity ($L_{\rm X} \sim
2$--$4 \times 10^{42} \ergps$) AGN in the sensitive DEEP2 X-ray data
that have similar host galaxies to those expected for the undetected
IR AGN. Indeed, IR AGN identified at lower redshift have relatively
low IR luminosities ($L_{\rm 4.5 \mu m} < 3 \times 10^{43} \ergps$),
which is, in general, below our sensitivity limit for IR AGN in
DEEP2. In turn, this may cause the factor of $\sim 2$ deficit found
between red-sequence DEEP2 X-ray and IR AGN observed in the normalized
populations (presented in Fig.~\ref{fig:histos_col}; see below).

\begin{figure*}[t]
\centering
\includegraphics[width=0.8\textwidth]{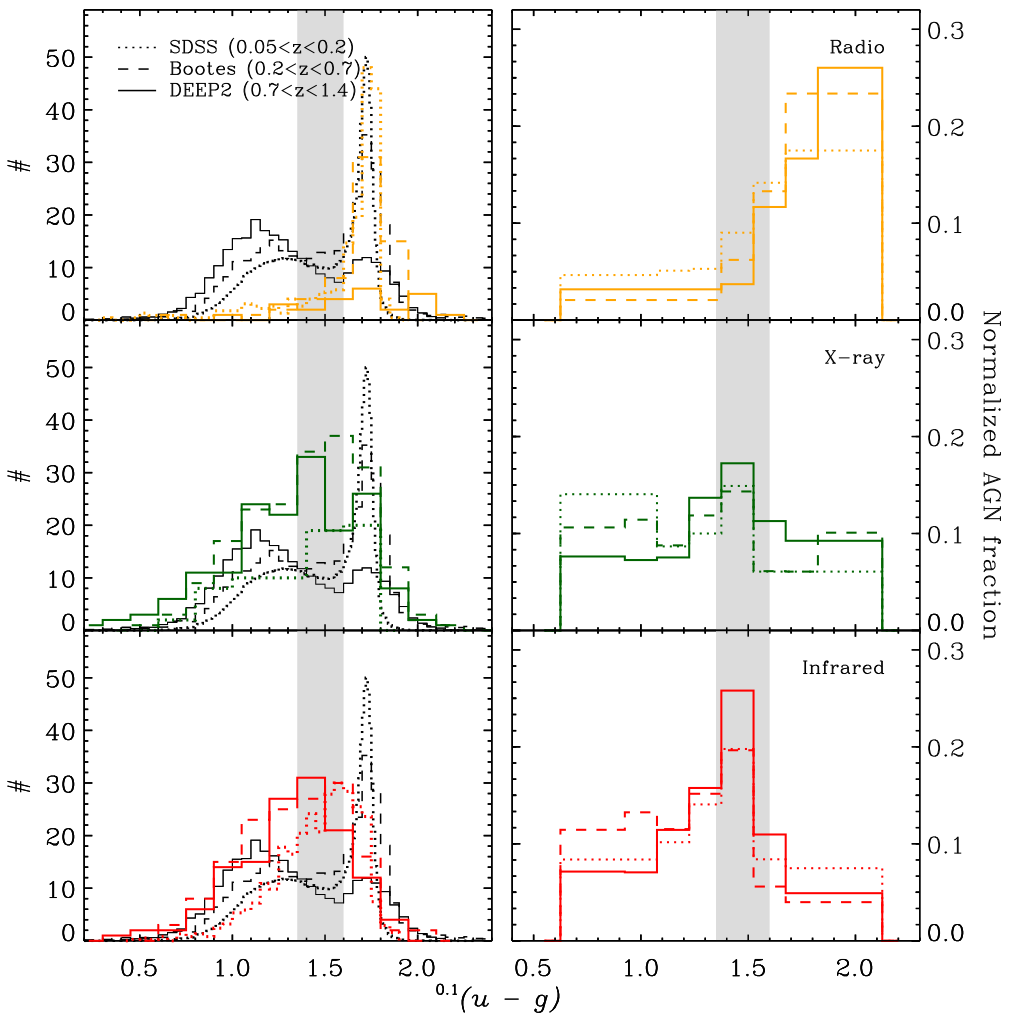}
\caption{{\bf Left:} Rest-frame $^{0.1}(u - g)$ color distributions,
  shifted to $z=0.1$ of radio (top panel; yellow), X-ray (middle
  panel; green) and IR (bottom panel; red) AGN. The respective parent
  galaxy populations in the SDSS ($0.05 < z < 0.2$; dotted line),
  \bootes ($0.2<z<0.7$; dashed line) and DEEP2 ($0.7<z<1.4$; solid
  line) surveys are shown in black, with the peak values scaled by a
  factor of 1/40 to match the AGN populations. The approximate
  position of the `green-valley' is shown in the shaded-region. {\bf
    Right:} Same as left panel with AGN histograms normalized to the
  respective parent galaxy populations.  AGN dominated by mechanical
  energy output are invariably hosted in red massive systems, and are
  anti-correlated with the hosts of radiatively efficient IR AGN;
  i.e., IR AGN are hosted in star-forming blue galaxies. When
  normalized for the underlying parent populations, we find little or
  no evidence for evolution in the host-galaxy color distributions
  within the last 9 Gyrs.}
\label{fig:histos_col}
\end{figure*}

Table~\ref{tbl:2d_ks} further shows that the cross-association between
radio AGN and X-ray/IR AGN consistently results in $KS(\rm MAP)<0.01$,
suggesting there is little or no link between the host galaxies of
luminous radio AGN and luminous X-ray/IR AGN. However, given that red
galaxies were marginally biased against in the original DEEP2
selection, this may be a selection effect brought about by the
distribution of the underlying galaxy
population. Fig.~\ref{fig:histos_col} shows the rest frame host galaxy
color distributions, for the three AGN selection wavebands. When
calculated as a fraction of the underlying galaxy population in each
survey, we confirm our K-S test results that after removing the effect
of the evolving galaxy population, the host galaxies of radio AGN
compared to IR/X-ray AGN are anti-correlated at all redshifts and this
does not appear to be an effect of the changing galaxy population. To
reduce small number statistical effects here, we use a bootstrap
analysis to randomly select AGN and galaxies from the main
populations. The area under the distributions are normalized to unity
to allow a better comparison between the AGN types and redshift
regimes.

Fig.~\ref{fig:histos_col} shows an over-density of IR AGN at
$^{0.1}(u-g) \sim 1.5$ throughout the three surveys, similar to that
seen for X-ray AGN in previous surveys (see
Section~\ref{subsec:col_mag}), which we marginally observe for X-ray
AGN here also. Such a peak in color is consistent with the expected
position of the `green-valley' (e.g.,
\citealt{nandra07,schawinski07b}) in the color--magnitude
diagram. There is also a factor $\sim$2 increase in the fraction of
radiatively-efficient AGN observed in blue-cloud galaxies over
red-sequence sources, suggesting that radiatively efficient AGN may,
in general, be confined to blue-cloud galaxies. Physically, this can
be interpreted as the AGN and star-formation being fueled by similar
mechanisms. This is unlikely to be ubiquitous as a population of X-ray
(and IR) luminous AGN are found in low-luminosity red-sequence
galaxies. It has been previously suggested that the effect of
host-galaxy dust-extinction may drive the identification of the
green-valley as an evolutionary staging ground for galaxies evolving
onto the red sequence (e.g., \citealt{cardamone10,bongiorno12}). Here
we show that the peak in the IR AGN population, combined with only a
small fraction of IR AGN existing in red-sequence galaxies, may
provide additional evidence that IR AGN are hosted in more obscured
galaxies and are not necessarily evolving directly from blue-cloud
disk-dominated galaxies into more passive elliptical systems.

When these results are taken together, after accounting for random
uncertainties and the changing underlying galaxy population, the
color-magnitude distribution of radio AGN host galaxies has remained
similar at all times in the last 9~Gyrs. While the distribution in
color and luminosity is substantially larger for X-ray and IR AGN,
there still appears to be little or no significant change in these
distributions within the epoch considered here. Put simply, we find
little evidence to suggest that the general host galaxy properties of
radio, X-ray, and IR AGN have changed or evolved in the last 9~Gyrs
(i.e., the host galaxy distribution for \{AGN$_k$\}$_{\rm R;X;IR}$ is
almost identical at each redshift slice).

There is now an emerging picture that points towards a distinct lack
of evolution in AGN hosts. At least in the X-ray AGN population, the
absence of significant evolution may even hold to $z\sim 3$ (e.g.,
\citealt{rosario13}). Furthermore, \cite{aird12} recently investigated
the color--stellar-mass distribution of X-ray detected AGN using
XMM-{\it Newton} and {\it Chandra} in the PRIMUS survey fields
\citep{coil11}, and found that while they observed evolution in the
specific fraction of galaxies that host AGN as a function of
redshift\footnote{Due to incompleteness effects, here we specifically
  do not investigate the absolute fraction of AGN in particular galaxy
  populations.}, the distribution and color dependence of the
stellar-mass independent AGN Eddington ratios (i.e., $\lambda_{\rm
  Edd} \propto L_{\rm X}/M_* $) have remained consistent since
$z\sim1$. Indeed, the similar distributions of X-ray AGN as function
of color and Eddington ratio found by \cite{aird12} may provide
further evidence that the AGN triggering mechanism and fueling source
are the same for radiatively efficient AGN. While a direct comparison
to the previous investigation is not possible due to the limited
coverage of the photometry in \bootes and DEEP2, which limits our
ability to place strong constraints on the stellar masses of
individual galaxies, we still find consistent and analogous results
with galaxy luminosity as an indirect proxy for the stellar mass. By
including the more sensitive and complete multi-wavelength data, we
have extended the previous results of \cite{aird12} to the lower
Eddington radio AGN population and to $z \sim 1.4$ for X-ray and IR
selected AGN, still finding no evolution in the overall AGN
population.

\subsection{Similar host-galaxy evolution of low-luminosity X-ray AGN}
\label{subsec:llxagn}

Unfortunately, source statistics are relatively poor for the X-ray AGN
identified in the SDSS. This is driven somewhat by the lack of a
dedicated homogeneous {\it Chandra} X-ray survey across the entire
SDSS region as well as the significant decrease in the average AGN
luminosity in the nearby Universe (i.e., AGN cosmic downsizing; e.g.,
\citealt{ueda03,barger05,hasinger05}). Furthermore, due to the
negative $K$-correction at X-ray energies, {\it Chandra} observations
at low redshift are more susceptible to nuclear extinction than
sources observed at higher redshift. Many galaxies with single X-ray
point sources coincident with the optical galaxy nucleus may contain
low luminosity or obscured AGN.

Fig.~\ref{fig:col_mag} additionally shows the color--magnitude
distribution of the 47 X-ray detected sources in the SDSS-DR7 with
$L_X \sim 10^{41}$--$10^{42} \ergps$, which also meet our
parent-galaxy selection methodology, outlined in the Appendix. This
lower X-ray luminosity AGN population has the same host galaxy
color--magnitude distribution as their more luminous
counterparts. Using the previous investigations of SDSS galaxies/AGN,
which analyzed local stellar and black hole mass distributions (e.g.,
\citealt{heckman04,kauffmann09}), and given the similar distribution
of host galaxies, it is likely that these AGN are low luminosity
and/or intrinsically obscured AGN. Hence, this population of $L_X \sim
10^{41}$--$10^{42} \ergps$ X-ray sources further reinforces our
observation of the similar host galaxy color--magnitude distributions
of X-ray sources across the entire redshift range. \footnote{To
  preserve our {\it clean} AGN selection, we do not include the $L_X
  \sim 10^{41}$--$10^{42} \ergps$ AGN population for any subsequent
  analyses, as the X-ray emission from a subset of these
  low-luminosity sources may still be dominated by stellar processes
  and X-ray binaries.}

There appears to be little systematic distinction in color--magnitude
space between the general galaxy population and either the low or high
X-ray luminosity AGN (i.e., X-ray AGN exist in all host galaxy
types). Such a lack of distinction between those galaxies that do and
do not host a radiatively efficient AGN may suggest that at some point
in cosmic time, the vast majority of galaxies have hosted an actively
growing black hole, and this is independent of the galaxy's individual
merger history, luminosity, stellar mass, or stellar
population. Whether a particular galaxy is observed to be growing the
central BH at a given luminosity is likely due to intrinsic
stochasticity of the accretion disk or variability of the obscuring
medium surrounding the central source (e.g.,
\citealt{lutz10,aird12,rosario13,shao13,hickox13}).

\begin{figure}[t]
\includegraphics[width=\linewidth]{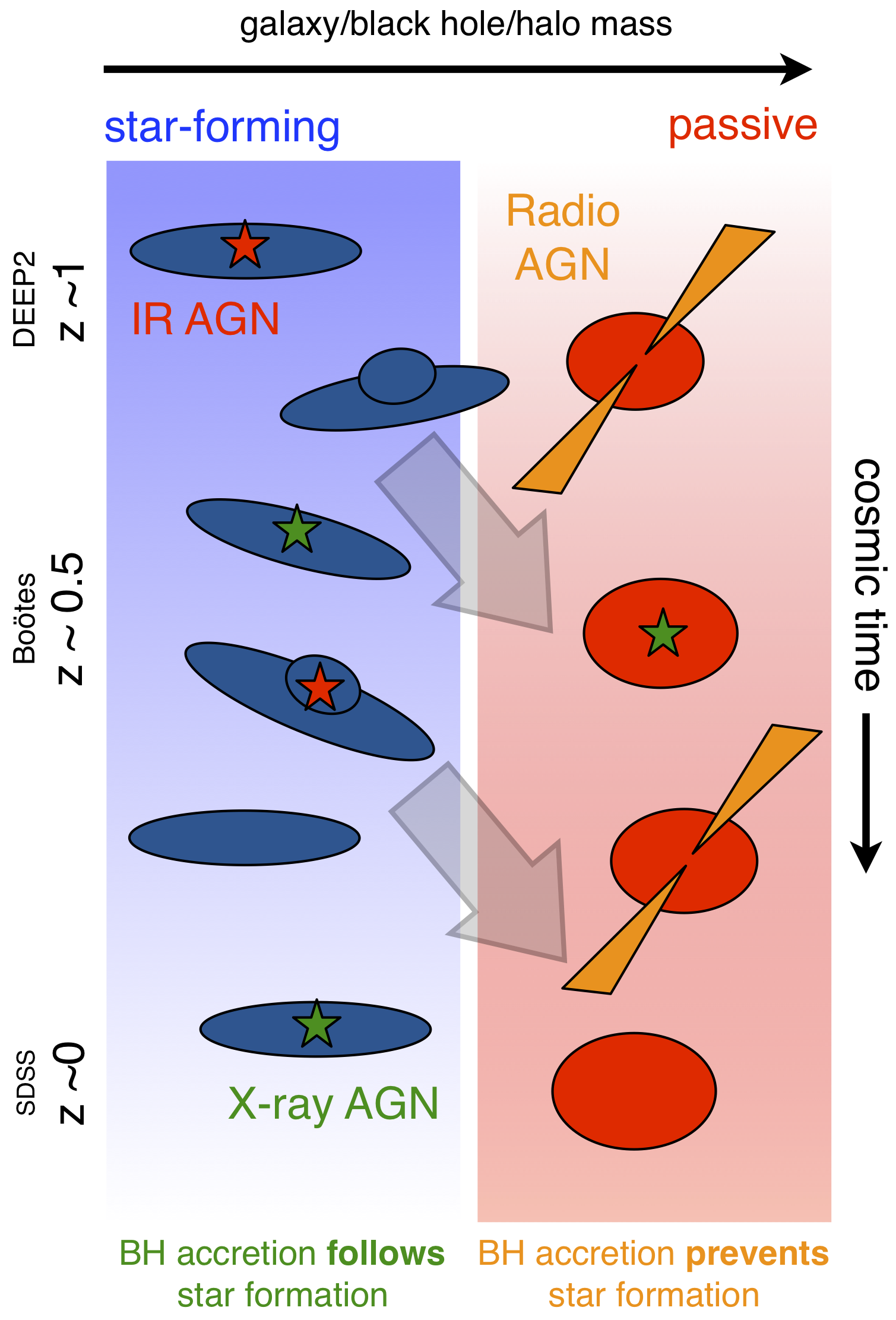}
\caption{Schematic diagram of the evolution of AGN host-galaxies
  throughout the last 9~Gyrs of Cosmic time. From the results
  presented here, we find that galaxy evolution follows an apparently
  similar path at all redshifts. All galaxies appear to begin as
  star-forming blue-cloud systems and end as passive red-sequence
  sources, once their dark matter halos have grown
  sufficiently. However, we show that those galaxies hosting IR, X-ray
  and/or radio AGN appear to follow a similar evolutionary path:
  radiatively-efficient rapid BH growth (IR/X-ray AGN) appears to be
  linked with those galaxies with large supplies of cool gas, while
  mechanically dominated (radio) accretion is associated with passive
  galaxies, which may also be responsible for preventing late
  star-formation.}
\label{fig:evol_schem}
\end{figure}

\subsection{AGN in dense groups \& cluster environments}

Galaxy populations in groups and clusters often differ in properties
from the field population. Galaxies in groups often consist of massive
spheroids hosting relatively old stellar populations, and these
galaxies likely occupy the high luminosity end of the red
sequence. Figs.~\ref{fig:col_mag} and \ref{fig:histos_col} show that
at all redshifts, luminous radio AGN with $P_{\rm 1.4.GHz} \gtrsim
10^{24}$W~Hz$^{-1}$ are tightly correlated with massive/luminous host
galaxies that lack strong on-going star-formation signatures, i.e.,
they have similar properties to those in dense group environments.

A substantial fraction of the red sequence build-up is expected to
occur between $0.2 < z < 1.4$ (e.g.,
\citealt{bell04,cooper06,faber07,moustakas13}).  Using the available
group catalogs for both \bootes (Vajgel et al., in prep.) and DEEP2
\citep{gerke12}, we can directly compare the environmental properties
of the AGN observed at or between these epochs. Both the \bootes and
DEEP2 group catalogs were constructed using similar Voronoi--Delaunay
methods. We identified the AGN hosted in red-sequence galaxies in both
surveys and compared these to the respective group catalogs. We found
consistent fractions of $\sim$26\% and $\sim$28\% IR and X-ray AGN
with red sequence hosts are group members, and these fractions remain
constant between $0.2 < z < 0.7$ and $0.7 < z < 1.4$ with no
significant evidence for evolution. Using the known positions of the
group barycenters, we assessed whether these AGN were also co-incident
(within 2~Mpc) of the group center. In general these are not at the
position of or even the galaxy closest to the barycenter. By contrast,
we find a suggestion of an increase in the fraction of radio AGN in
groups between \bootes and DEEP2. In DEEP2, at $0.7 < z < 1.4$, only
$\sim$19$\pm10$\% (3/16) of the radio AGN are group members,
consistent with the fraction found for X-ray and IR AGN in DEEP2;
while $\sim$40$\pm9$\% (19/48) of radio AGN are in groups by $0.2 < z
< 0.7$ in \bootes, and these AGN are predominantly consistent with the
group center. These results suggest the location/environment of the
(central) galaxy within the group may drive the production of the
radio jet.

At $z \sim 1$, group/cluster environments are relatively rare and have
not yet become fully established with warm/hot gas halos. Hence, at $z
\sim 1$, we observe only a small fraction ($\sim 20$\%) of radio
galaxies are in group environments. The remaining DEEP2 radio galaxies
are additionally IR and X-ray detected sources, these are more likely
linked with HEX radio activity. By contrast, at $z \sim 0.5$, the
group environments are in place. The warm intracluster gas inhibits
efficient accretion, creating powerful radio jets. These radio jets
and subsequent radio lobes have the effect of preventing further
cooling of the group gas, and restrict the production of new
stars. This picture can be further substantiated as only two of the 13
radio galaxies present in groups at $0.2 < z < 0.7$ are also
identified as AGN at X-ray and IR wavelengths. Furthermore, it appears
that radio galaxies follow the build-up of group environments in
red-sequence galaxies. It is then the existence of the powerful radio
lobes, which are injecting mechanical energy back into the
group/cluster that prevents efficient gas cooling and restricts star
formation, sustaining the radio AGN host galaxy as a red-sequence
system (so-called `maintenance-mode' feedback;
\citealt{best05b,rafferty06,mcnamara07,kauffmann08,jones10,bower12}.

\section{Discussion} \label{sec:discussion}


Observational studies of low and high-redshift radio galaxies have
found evidence for large-scale mechanically driven outflows in the
form of (often) megaparsec-scale relativistic radio jets (and
associated lobes) originating from the AGN. Given that such radio jets
are almost exclusively observed in central massive `red and dead'
systems (e.g., \citealt{best05b, wake08, hickox09}), it has been
suggested that a quenching phase that transitions galaxies from having
large cool gas reservoirs to being dominated by virialized hot
atmospheres may be essential for the build-up of the red sequence of
galaxies \citep{cooper06,bower06,croton06,bower08}.

In light of this, the current consensus is that galaxies likely begin
as star-forming systems and become relatively passive once their dark
matter halos reach sufficient mass and/or they undergo freefall into a
second large halo i.e., they experience tidal-stripping due to
interaction with a group or cluster. We have found that powerful radio
galaxies tend to be found in dense gas-rich environments located near
the group barycenter and this appears to be true at all redshifts (to
$z<1.4$). These radio AGN are likely providing the heating source,
through maintenance-mode feedback, to have a profound effect on the
gas present in the host galaxy and by extension the gas within the
dark matter halo. This AGN feedback is inhibiting subsequent gas
cooling and the rapid production of new stars. Indeed, as the fraction
of galaxies hosting radio sources is a strong function of both black
hole and stellar mass (e.g., \citealt{best05b}), it would appear that
powerful radio sources closely follow the build-up of the most massive
and luminous systems on the red sequence.

By contrast to the radio AGN population, recent large scale galaxy
population studies, focusing on the coincident presence of rapid star
formation activity and radiatively-efficient AGN recognized as X-ray
and/or IR luminous central sources, have revealed contradictory
results. In spite of the strong belief that mergers and the presence
of an AGN play a significant role in quenching the on-going star
formation and transitioning galaxies from the blue cloud to the red
sequence, many recent studies now suggest that X-ray/IR luminous AGN
have not played a substantial role in the build-up of the red
sequence. The prefered picture is secular galaxy evolution (e.g.,
\citealt{kocevski12,schawinski12}). However, even for a galaxy that is
evolving through secular processes, the presence of a rapidly growing
BH still necessitates a large cool gas supply to fuel accretion. When
combined with a correlation between high-accretion rate AGN and star
formation (e.g., \citealt{lutz08,bonfield11,symeonidis11,juneau13}),
this has led to the hypothesis that these two processes may be fueled
from the same gas reservoirs (e.g., \citealt{chen13,hickox13}). As
part of our K-S evaluation (see Section~\ref{subsec:hostgal_evol}) of
AGN populations at $0.05 < z < 1.4$, we found that the bulk of the
radiatively efficient (X-ray; IR detected) AGN are in predominantly
star-forming (blue cloud) galaxies. Taken together, these results
provide evidence for a broad relationship between the coincident
presence of an AGN and ongoing star formation.  In turn, this may lead
to a form of self-regulation for the growth of the central BH. We also
find that any trends with redshift observed for the galaxy color, and
by extension the integrated star formation of the host, appear to be
similar for galaxies hosting and not hosting AGN, suggesting that
while the AGN and on-going star formation may be coincident and fueled
simultaneously, the instantaneous presence of the
radiatively-efficient (X-ray or IR detected) AGN is having little
effect on the evolution of the host galaxy in the last 9
Gyrs. Furthermore, and as we noted previously, there is also a
population of AGN present in weakly star-forming systems. Taken
together, it therefore appears that AGN and galaxy evolution may
follow similar paths at all redshifts (see Fig.~\ref{fig:histos_col}).

Figure~\ref{fig:evol_schem} presents a schematic diagram to represent
the galaxy colors/types which host AGN throughout the last 9~Gyrs of
Cosmic time. At all redshifts, mechanically dominated (low excitation)
radio AGN are hosted in a separate galaxy population from the
radiatively efficient X-ray/IR AGN. This can be understood by powerful
radio galaxies requiring a massive host with a large BH, an old
stellar population, and possibly located near the center of a large
group or cluster environment. By contrast, the radiatively efficient
IR and X-ray AGN are generally ($> 80\%$) hosted in lower mass
galaxies hosting smaller black holes with ample cool gas to fuel
higher accretion. Those AGN hosted in galaxies with redder blue-cloud
colors are likely to be dust rich and/or have smaller reservoirs of
cool gas to fuel young star formation and the central BH.

Finally, the typical star formation gas depletion time scale (i.e.,
the time taken to convert the available cool gas supply into stars) is
of the order several gigayears for an individual massive spiral
galaxy. Combining this with the strong build up of intermediate mass
galaxies on the red sequence since $z\sim1$ (e.g.,
\citealt{faber07,moustakas13}), passive evolution alone is not
sufficient to explain the rapid change in the galaxy luminosity and
mass function in the last 7~Gyrs (e.g., Marchesini et al. 2009). The
disruption to the host-galaxy gas, preventing late star formation
activity, has long since been attributed to the presence of a central
AGN as the required quenching mechanism. Hence, the lack of evidence
for evolution in the host-galaxy colors of the radiatively-efficient,
X-ray and IR AGN presents a problem for the requirement of the rapid
build-up of red-sequence galaxies. This likely presents two
evolutionary paths: (1) the presence of an AGN is having little or no
impact on the evolution of galaxies, and a different mechanism is
quenching the star formation in field galaxies (e.g., increased
supernovae activity; self supression through starburst winds); or (2)
the stochastic presence of a variable (and at times extremely
luminous) AGN has a short-lived but significant effect on the host
properties (e.g.,
\citealt{schawinski10,novak11,sarajedini11,gabor13}). Such an event
would be similar to a spectacular quasar-phase of galaxy evolution.
Such sources are outside our sample because we require accurate
measurements of the host galaxy, which is outshone when in the
presence of a quasar.

\section{Summary \& Conclusions} \label{sec:conclusions}

We have presented a statistical analysis of the host galaxy rest-frame
colors and luminosities of AGN at $0<z<1.4$ that were identified at
radio, X-ray and IR wavelengths within the SDSS, Bo\"{o}tes, and DEEP2
surveys. Using the available galaxy catalogs within the these three
fields, we constructed a parent sample of 330,811 galaxies that are
matched for luminosity and galaxy color, and corrected for passive
evolution to allow a direct comparison of the sources identified
throughout the $0<z<1.4$ epoch. We used the available multi-wavelength
ancillary data from WISE, {\it Spitzer}-IRAC, {\it Chandra}-ACIS
Westerbork radio and VLA to identify those galaxies hosting radio,
X-ray and IR AGN. Using this large multi-wavelength homogeneous galaxy
sample, we presented the color--magnitude distributions for the main
galaxy sample and the AGN host galaxies. Through an MCMC 2-D K-S
analysis, we found that at all redshifts, radio AGN with $P_{\rm 1.4
  GHz} > 10^{24}$~WHz$^{-1}$, powered by advection-dominated accretion
mechanisms, are systematically hosted in luminous red-sequence
galaxies. Conversely, and in line with previous studies, X-ray (with
$L_{\rm X} > 10^{42} \ergps$) and IR AGN (selected through IR
color--color techniques), which are undergoing radiatively-efficient
accretion are, in general, hosted in a separate population of blue
cloud galaxies and lower luminosity red galaxies. Despite the build up
of the red sequence of galaxies throughout the epoch considered here,
we find little or no statistical evidence that those galaxies that are
actively growing their central black holes have evolved separately
from the galaxies not hosting AGN. However, once a galaxy has become
established on the red sequence, the coincident presence of a central
radio source may inhibit subsequent star formation in these systems
and prevent them from returning to the blue cloud. Hence, based on the
statistical analyses presented here, AGN hosts appear to follow
similar evolutionary paths to the main galaxy population at all times
in the last 9~Gyrs.

\acknowledgments

We are grateful to J. Donley for kindly providing a set of
star-forming infrared spectral energy distribution templates and
associated photometry measurements. We also acknowledge enlightening
conversations with W. Joye. This publication makes use of data
products from the Wide-field Infrared Survey Explorer, which is a
joint project of the University of California, Los Angeles, and the
Jet Propulsion Laboratory/California Institute of Technology, funded
by the National Aeronautics and Space Administration. This work is
also based in part on observations made with the Spitzer Space
Telescope, which is operated by the Jet Propulsion Laboratory,
California Institute of Technology under a contract with
NASA. Furthermore, we make use of data from the Sloan Digital Sky
Survey, which is managed by by the Astrophysical Research Consortium
for the Participating Institutions and funding provided by the Alfred
P. Sloan Foundation, the Participating Institutions, the National
Science Foundation, the U.S. Department of Energy, the National
Aeronautics and Space Administration, the Japanese Monbukagakusho, the
Max Planck Society, and the Higher Education Funding Council for
England. Finally, we acknowledge use of NASA's astrophysics data
system.

{\it Facilities:} \facility{CXO (ACIS); Spitzer (IRAC), WISE}.

\bibliography{./bibtex1}


\newpage
\appendix

\section{Galaxy Sample Description \& Selection}
\label{appen:sample}

In this appendix material we present an outline of the survey fields
from which our main galaxy sample is selected from and the composition
of the multi-wavelength ancillary data present within the
fields. Furthermore, we include an in-depth description of the
like-for-like galaxy matching procedure between the three surveys and
redshift slices, along with the appropriate optical $K$-correction
procedure to calculate rest-frame colors and galaxy luminosities.

\subsection{The Sloan Digital Sky Survey}
\label{sec:sdss}

Our sample of relatively low-redshift galaxies, presented here, is
selected from the seventh data release of the Sloan Digital Sky Survey
(SDSS-DR7; \citealt{sdss_dr7}). The SDSS-DR7 is currently the largest
publicly available optical extragalactic spectroscopic catalogue
($\approx 8032$~deg$^{2}$) containing $\sim$900,000 spectroscopic
galaxy redshifts. The sources are primarily identified in the northern
Galactic sky with photometry determined from 5-band $ugriz$ filters at
$\lambda \sim 3550$--8900\AA. The spectroscopic catalog is complete to
$r<17.77$ magnitudes. Full optical data products are available
directly from the SDSS archive, but to ensure the accuracy of the
emission line identifications and photometric measurements, here we
adopt the dedicated reprocessed data products from the MPA-JHU DR7
catalog (e.g.,
\citealt{brinchmann04,kauffmann03c,tremonti04,lamassa13}). The MPA-JHU
catalog contains photometric and spectral measurements for $\sim
800,000$ unique galaxies. For our purposes, we select the 592,300
spectroscopically targeted galaxies in the redshift range $z \sim
0.05$--0.2.

With the advent of all-sky surveys and homogeneous data reduction
systems, large moderate-sensitivity multi-wavelength observations do
now exist which cover substantial fractions of the SDSS sky region. At
radio wavelengths, with $\sim$950,000 sources detected down to 1 mJy,
the Very-Large-Array Faint Images of the Radio Sky at
Twenty-Centimeters (hereafter, VLA-FIRST; \citealt{becker94,becker95})
survey covers $\sim$10,600deg$^2$ of the sky, with $\sim$94\% coverage
of the SDSS region. At mid-infrared (mid-IR) wavelengths, we adopt the
all-sky catalog of the Wide-field Infrared Survey Explorer (hereafter,
WISE) to provide moderate spatial resolution 4-band IR photometry for
the sources identified across the SDSS region \citep{wright10}. We
describe our source-matching procedure between WISE and SDSS as well
as our AGN identification technique in Section 3.1. While dedicated
{\it Chandra} X-ray surveys exist for both the \bootes and DEEP2
regions, a complete {\it Chandra} survey of the full SDSS region is
prohibitively expensive. However, archival {\it Chandra} data already
exist for $\sim$130~deg$^2$ (after accounting for overlapping
observation regions) between SDSS and {\it Chandra} as part of the
Chandra Source Catalog. After removal of target {\it Chandra} sources,
this provides a representative sample of SDSS sources with sensitive
X-ray observations that are comparable to the \bootes and DEEP2
surveys.

\subsection{The NDWFS Bo\"{o}tes field}
\label{sec:bootes}

With $\sim 41,000$ identified galaxies, primarily at $z\sim0.2$--0.7,
the 9.3~deg$^2$ NOAO Deep Wide-Field Survey (NDWFS;
\citealt{jannuzi99}) \bootes field with homogeneous spectroscopic
coverage from the AGN and Galaxy Evolution Survey (AGES;
\citealt{kochanek12}) provides an ideal survey area to probe the
moderate redshift regime. The \bootes field (RA=14h32m00s,
Dec=31d16m47s [J2000]) lies at both high galactic and ecliptic
latitude, and hence, is relatively free of significant foreground
contamination. Optical photometry for the full NDWFS \bootes field was
conducted with the 4m Mayall Telescope at the Kitt Peak National
Observatory in $B_{\rm W}$, $R$ and $I$ filters with $z$-band
photometry for $\sim 62$\% of the survey.  The AGES redshift survey is
complete to $I<20.5$~magnitudes, with redshifts determined from MMTO
Hectospec observations \citep{fabricant05}, with $R\sim1000$ in the
wavelength range $4500 < \lambda < 8900$\AA, and covers $\sim
7.74$~deg$^2$ of the NDWFS \bootes area. Furthermore, contiguous
multi-wavelength observations in the mid-infrared ({\it Spitzer}-IRAC
and MIPS; \citealt{eisenhardt04,stern05,ashby09}), far-infrared ({\it
  Herschel}-PACS and SPIRE), radio (Westerbork 1.4GHz;
\citealt{devries02}) and shallow X-ray ({\it Chandra}-ACIS;
\citealt{murray05,kenter05,brand06}) exist across the AGES region.

\subsection{The DEEP2 Galaxy Redshift Survey}
\label{sec:deep2}

The $3.6$~deg$^2$ DEEP2 Galaxy Redshift Survey
\citep{davis03,newman13}, is currently the widest area deep
spectroscopic survey of $z \sim 1$ galaxies, making it an ideal survey
to target large numbers of AGN to $z \sim 1.4$. The photometric
catalog for DEEP2 \citep{coil04} is derived from Canada-France-Hawaii
Telescope (CFHT) images from the 12x8k mosaic camera
\citep{cuillandre01} in $B$, $R$, and $I$ filters across four
non-contiguous ($\sim$0.6--0.9~deg$^2$) sky regions and is complete to
$R < 25.2$~magnitudes. The fourth data release of the spectroscopic
survey contains the spectra for $\sim 50,300$ galaxies
(\citealt{newman13}; with $R_{\rm AB} < 24.1$). These galaxies are
primarily in the redshift range $z \sim 0.7$--1.4 with spectroscopic
redshifts determined from data collected using the DEIMOS spectrograph
($R \sim 5000$ in the wavelength range $6400 < \lambda < 9200$\AA) on
the Keck II telescope. All four DEEP2 fields have X-ray coverage using
the Chandra X-ray Telescope (the XDEEP2 survey; \citealt{goulding12b})
with combined exposures in the range $\sim 10$--800ks. Field~1 of
DEEP2 has dedicated VLA radio \citep{ivison07,willner12} and Spitzer
mid-IR photometry \citep{barmby08} as part of the All-wavelength
Extended Groth Strip International Survey (AEGIS; see
\citealt{davis07} for an overview). DEEP2 Fields 2 and 4 have
sensitive Spitzer mid-IR photometry (see Section~\ref{sec:AGN} of the
main manuscript). Finally, all four fields are covered at radio
wavelengths by the VLA as part of the FIRST survey.

\subsection{A color--luminosity-matched galaxy sample}
\label{sec:sample_match}

The three surveys considered here, i.e, SDSS, \bootes and DEEP2, have
independent photometric detection and spectroscopic selection
criteria. For those survey regions with complete and homogeneous
spectroscopic coverage, the limiting source magnitude in the SDSS,
\bootes and DEEP2 surveys is $r < 17.77$, $I < 20.5$ and $R<24.1$ AB
magnitudes, respectively. While these surveys select sources at
roughly the same observed wavelengths, albeit with varying degrees of
sensitivities, the large ranges in target redshifts result in the
rest-frame selection being markedly different. In particular, the
rest-frame UV selection of galaxies in DEEP2 results in good
senstivity towards blue star-forming galaxies, which is further
enhanced by the ability to detect strong [O{\sc ii}]$\lambda$3727 at
$z<1.4$. By contrast, this results in a loss of sensitivity towards
red galaxies with smiliar masses to the blue sources that would
otherwise be detected in DEEP2 at the same redshift, this is due
mainly to changes in the overall spectral shape of the dominant
stellar population between restframe $B$ and $R$-bands. Such selection
effects have been previously investigated by \cite{willmer06} and
\cite{faber07}. Conversely to the DEEP2 selection, the rest-frame red
optical selection of the SDSS survey will result in a marginally
reduced sensitivity towards nearby faint blue galaxies and greater
completeness towards red galaxies. Hence, to carry out a comparison of
the AGN host galaxy properties, these optical selection effects must
be mitigated. Here, we build on the previous methodology of
\cite{blanton06} and describe our technique to perform $K$-corrections
to the observed photometry and match the source catalogs based on
their rest-frame optical measurements.

\subsubsection{Galaxy {\it K}-corrections}
\label{subsec:gal_kcorr}

We begin by performing standard optical $K$-corrections independently
across the three surveys. For the \bootes (NDWFS) photometry, we
convert Vega magnitudes to AB using corrections of: $B_W=$ +0.02 mag,
$R=$+0.19 mag, $I=$+0.44 mag (e.g., \citealt{hickox09}). We also
correct the SDSS and \bootes photometry for interstellar extinction
\citep{schlegel98}. The photometry for DEEP2 is provided
extinction-corrected in AB standard \citep{coil04}.

For this investigation, we are particularly interested in constraining
the host galaxy properties. Hence, we select all sources in \bootes
and DEEP2 with $P_{\rm gal} > 0.9$ (defined in \citealt{coil04}) to
robustly remove stars and quasars (i.e., extremely luminous AGN that
dominate the galaxy starlight). In the SDSS, sources are selected from
the SDSS-DR7 galaxy catalog, which by design does not include Galactic
sources or quasars (see \citealt{strauss02,richards02}). Photometric
$K$-corrections are performed using the publicly-available {\tt C} and
{\tt IDL} tool, {\sc kcorrect} (v4.2)\footnote{The current version of
  {\sc kcorrect} is available at
  \url{http://howdy.physics.nyu.edu/index.php/Kcorrect}}. {\sc
  kcorrect} fits a carefully chosen set of stellar population
synthesis (e.g., \citealt{bruzual03}) and photoionization/shock (e.g.,
MAPPINGS-III; Sutherland \& Dopita 1993) templates to the observed
photometry, where the templates have been shifted to the known
redshift of the source. In a standard implementation of {\sc
  kcorrect}, the best-fit template is then deprojected at rest-frame
to calculate the required $K$-correction for the survey photometric
filters and/or produce the intrinsic photometry in custom filters
defined for arbitrary redshifts. Such measurements can then be used to
construct rest-frame color--magnitude diagrams for large source
populations. However, here, we must first consider the inherent source
selection bias in the SDSS, \bootes and DEEP2 surveys, caused by the
differing median source redshifts and the observed-frame selection
techniques.

\begin{figure}[t]
\centering
\includegraphics[width=0.5\linewidth]{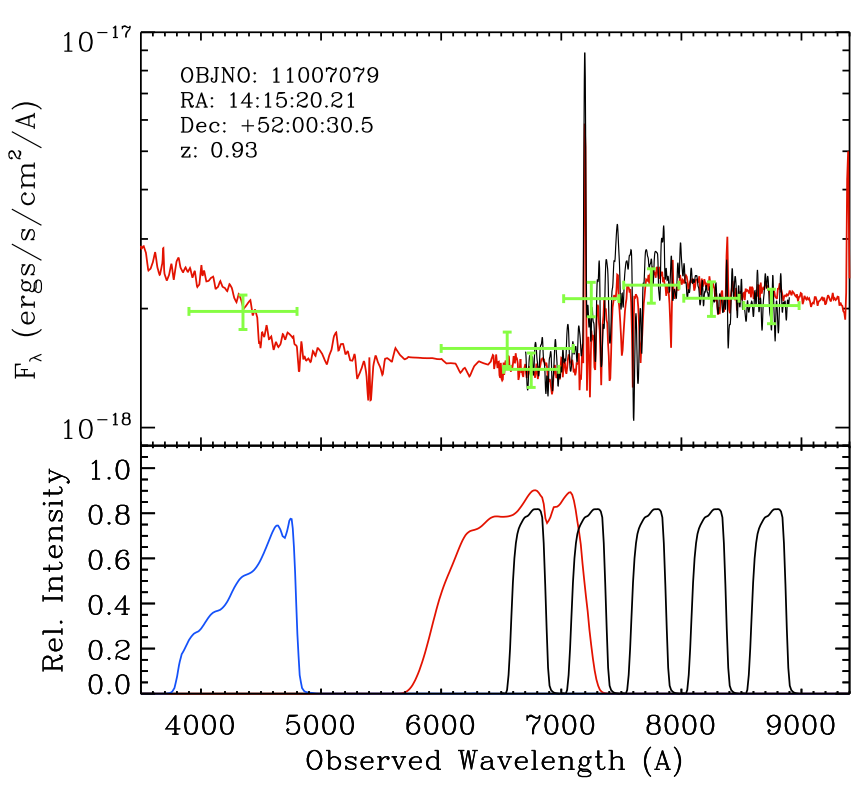}
\caption{{\it Top panel:} Example of a typical {\sc kcorrect} fit to a
  DEEP2 galaxy using the measured $BR + I'_{1:5}$ photometry (green
  points). Flux density as a function of observed wavelength is shown
  for the DEIMOS spectroscopy (black; Newman et al. 2013),
  stellar$+$photoionization$+A_{\rm V}$ template produced by {\sc
    kcorrect} is shown with red solid line. {\it Lower panel:}
  Photometric responses for the standard DEEP2 $B$ (blue) and $R$
  (red) filters, and our custom narrow $I'$ filters (black) used
  during the {\sc kcorrect} fitting procedure.}
\label{fig:sample_spec}
\end{figure}

Sources identified in the spatial regions covered by \bootes and DEEP2
are observed with relatively similar $BRI$ photometric
filters. However, given the higher median redshift of the DEEP2 sample
($z \sim 0.87$) compared to that of \bootes ($z \sim 0.34$), this
translates to sampling shorter and narrower rest-frame wavelengths for
galaxies in DEEP2, particularly for the highest redshift sources. In
principle, $K$-corrections derived for sources with 3-band photometry
covering only $\lambda \sim 2290$--4210\AA~in the rest-frame (similar
to the wavelength range covered by NUV$_{\rm GALEX}$--U$_{\rm
  Johnson}$) may fail to fit the majority of the reprocessed light due
to foreground extinction, as well as failing to account for the
starlight from old stellar populations which typically emits at longer
optical wavelengths. To partially mitigate these effects, we
artificially over-sampled and extended the photometric measurements to
$\lambda \gtrsim 9000$\AA\ using the available DEIMOS spectroscopy.

For each DEEP2 galaxy we predicted synthetic photometry in five
pre-defined medium-band filters to supplement the standard $BRI$ data
during the {\sc kcorrect} fitting process. We constructed five
relatively narrow (500\AA) filters that are based on the response
shape of the $i_{\rm SDSS}$ filter, and centered at $\lambda_{0}
=$\{6750,7250,7750,8250,8750\AA \}. The DEIMOS spectra were convolved
with the $I$-band filter and flux-calibrated using the I-band
measurement after applying a simple aperture correction between the
DEIMOS slit and the photometric aperture at the effective wavelength
of the filter ($\lambda_{\rm eff} \sim 8060$\AA).  The individual
source spectra were then convolved with the five constructed
$I'_{1:5}$ filters, and synthetic photometric measurements were
extracted. We note that the choice and use of five synthetic
photometric points is somewhat arbitrary. However, we found that five
additional photometric measurements can adequately map the depth of
the $D_{4000}$ break (visible at $z \sim 0.7$--1.4) in the galaxy
spectra, which is primarily used to determine the stellar population
contributions in synthesis modelling. We tested the use of five
additional filters (for a total of seven filters) versus six and seven
additional filters, and while we found that increasing numbers of
photometric points can map the $D_{4000}$ break and emission-line
features with greater accuracy, the overall improvement to the
$\chi$-squared value was marginal. Due to the large number of spectra
to be analyzed ($\sim$50,000), we found that five additional
narrow-band photometric points was an adequate trade-off between
increasing accuracy and processing time.

$K$-corrections were calculated for the DEEP2 galaxies using the
method outlined above and the $BRI'_{1:5}$ photometric
measurements. In Fig.~\ref{fig:sample_spec}, we show an example {\sc
  kcorrect} fit to the $BRI'_{1:5}$ photometry for a typical
$z\sim0.9$ DEEP2 galaxy. The fitted {\sc kcorrect} template adequately
reproduces the main features observed in the DEIMOS spectra, giving
strong confidence that the calculated rest-frame photometry is
appropriate for our purposes.

\subsubsection{Passive evolution corrections \& selection matching}
\label{subsec:passive_correct}

Using the best-fit rest-frame galaxy/AGN template determined for each
source, we predicted if the source is sufficiently luminous at $z_0$
to be detected in the other two survey fields (e.g., is a particular
SDSS galaxy sufficiently bright that it could have been detected in
the \bootes and DEEP2 bandpasses). For each source, we use a
Metropolis-Hastings sampler to randomly select two redshifts, $z_1$,
$z_2$ from the appropriate survey redshift distributions (i.e., from
the two surveys that the source was not originally selected in) to be
applied to the best-fit rest-frame template. We applied
passive-evolution corrections to the source luminosity to shift the
observed-frame {\sc kcorrect}-template at $z_0$ to $z_1$ and
$z_2$. For those galaxies lying in the blue-cloud of the
color-magnitude diagram, we evolved the templates by $+$0.98
magnitudes per unit redshift, and for red sequence sources by$+$0.66
magnitudes per unit redshift \citep{blanton06}. The larger correction
for the blue-cloud galaxies simultaneously includes a
passive-evolution correction as well as accounting for the overall
decline in the specific star-formation rate in the last 9~Gyrs. We
also note here, that the use of a pseudo-random redshift for each
source selected from the survey redshift distribution, as opposed to
the redshift-limit of the survey, ensures a meaningful and realistic
luminosity distribution for the galaxy population. Redshifted and
luminosity corrected source templates are then deprojected back into
the observed-frame photometric filters for the survey bandpasses, and
those sources that meet the selection criteria for all three surveys
are included in the final matched source sample.

For the galaxies identified in DEEP2, we found that the introduction
of the synthetic $I'_{1:5}$ photometry, as part of the {\sc kcorrect}
analysis (see Section~\ref{sec:sample_match}), significantly improved
the quality of the template fitting, resulting in reduced
uncertainties (on average by $\sim$85\% from the 3-band DEEP2 data) in
the inferred $^{0.1}u$ and $^{0.1}g$ band photometry. Furthermore, we
observed that the wide spread in $^{0.1}$($u$--$g$) color
($\sim$0.5~mags) found for passive red-sequence DEEP2 galaxies (see
Section~\ref{subsec:col_mag}) was not significantly decreased compared
to using a single $I$-band photometric point. This suggests that other
systematic effects may dominate the {\sc kcorrect} fitting process
and/or the red-sequence is intrinsically broader in $^{0.1}$($u$--$g$)
color at higher-redshift. However, further analysis of the intrinsic
color--magnitude distribution at $z \sim 1$ is beyond the scope of
this investigation, and is arguably not suited to only the three-band
$BRI$ photometry available in DEEP2.

Finally, we note that we have not attempted to correct for potential
biases resulting from the requirement of a spectroscopic redshift for
each source. For example, galaxies lacking emission lines or other
obvious emission features that would be used to discern a redshift
will be uniformly selected against across our entire galaxy
sample. Given that we have tailored the galaxy selection to the
redshift regimes where the individual surveys are most sensitive, this
bias will most likely affect the low-luminosity end of each sample and
not the galaxies hosting luminous AGN. Furthermore, while less of a
potential issue due to repeat exposures and observations
shifting/dithering, galaxies with very close-by neighbors as projected
onto the sky may be systematically avoided or result in redshift
failures due to blending effects.

\end{document}